\documentclass{iopart}

\usepackage[OT1,T1]{fontenc} 
\usepackage{pstricks,pst-node,pst-text,pst-3d,multirow,bigstrut}
\usepackage{psfrag,epsfig}
\usepackage{graphics}
\usepackage{psfig}

\graphicspath{{./fig/}}

\def\FC{{\mathcal F}}

\def\HC{{\mathcal H}}
 
\def\LC{{\mathcal L}}

\def\TC{{\mathcal T}}

\def\vect#1{\mbox{\boldmath{$#1$}}}
\def\evect#1{\mbox{\boldmath{$e$}}_{#1}}

\def\ra{\rightarrow}

\font\tenbb=bbold10 scaled \magstep1

\font\sevenbb=bbold7 scaled \magstep1

\font\fivebb=bbold5 scaled \magstep1

\newfam\bbfam
\textfont\bbfam=\tenbb
\scriptfont\bbfam=\sevenbb
\scriptscriptfont\bbfam=\fivebb
\def\bb{\fam\bbfam\tenbb}

\def\Rb{\mbox{$\bb R$}}

\def\Pb{\mbox{$\bb P$}}

\def\1b{\mbox{$\bb 1$}}

\begin{document}

\noindent {\sc Preprint} \hfill  \today

\title{Flip dynamics in three-dimensional random tilings}

\author{V. Desoutter, N. Destainville \medskip \\
\rm Laboratoire de Physique Th\'eorique~-- IRSAMC \\
UMR 5152 CNRS/Universit\'e Paul Sabatier, \\ 
118, route de Narbonne, 31062 Toulouse Cedex 04, France.}

\begin{abstract} 
We study single-flip dynamics in sets of three-dimensional rhombus
tilings with fixed polyhedral boundaries. This dynamics is likely to
be slowed down by so-called ``cycles'': such structures arise when
tilings are encoded {\em via} the ``partition-on-tiling'' method and
are susceptible to break connectivity by flips or at least ergodicity,
because they locally suppress a significant amount of flip degrees of
freedom. We first address the so-far open question of the connectivity
of tiling sets by elementary flips. We prove exactly that sets of
tilings of codimension one and two are connected for any dimension and
tiling size. For higher-codimension tilings of dimension 3, the answer
depends on the precise choice of the edge orientations, which is a
non-trivial issue.  In most cases, we can prove connectivity despite
the existence of cycles. In the few remaining cases, among which the
icosahedral symmetry, the question remains open. We also study
numerically flip-assisted diffusion to explore the possible effects of
the previously mentioned cycles. Cycles do not seem to slow down
significantly the dynamics, at least as far as self-diffusion is
concerned.
\end{abstract}



\bigskip

\section{Introduction}

Rhombus tilings in dimensions 2 and 3 have been an interdisciplinary
subject of intensive study in the two last decades, both in
theoretical solid state physics because of their strong relation with
quasicrystals~\cite{levine84,elser85,henley91}, as well as in
theoretical computer science or in more fundamental
mathematics~\cite{penrose74,kenyon93,elnitsky97,bailey97,cohn98,cohn01,Linde01}.
Rhombus tilings are coverings of a portion of Euclidean space, without
gaps or overlaps, by rhombi in dimension two and rhombohedra in
dimension three. The ``cut-and-project'' process is a standard
method~\cite{elser85,cutandproject} to generate such tilings. It
consists in selecting sites and tiles in a $D$-dimensional cubic
lattice and in projecting them onto a $d$-dimensional subspace with
$D>d$; $d$ is the dimension of the tilings and the difference $D-d$ is
usually called their {\em codimension}. The class of symmetry of a
tiling is related to $D$ and $d$ and such tilings will be denoted by
$D \ra d$ tilings.  Icosahedral tilings that are widely studied in
quasycristal science, are $6 \ra 3$ tilings of codimension 3. We
consider in this paper three-dimensional tilings of codimensions
ranging from 1 to 4.  We also make incursions in the general $D \ra d$
case whenever it is possible.

The ``generalized partition'' method used in this paper to generate
tilings is a variant of the previous one which has proven useful in
several circumstances to manipulate and count
them~\cite{elser84,mosseri93,mosseri93B,destain97,bailey97,octo01,arctic}.
The principles of the method will be recalled below. However, we
should already mention that the tilings generated by this technique
have specific fixed boundaries. They are polygons in dimension 2 and
polyhedra in dimension 3, all of them belonging to the class of {\em
zonotopes}~\cite{coxeter}. 

As compared to perfect quasiperiodic rhombus tilings, such as the
celebrated Penrose tilings~\cite{penrose74}, the so-called ``random
tilings''~\cite{henley91} have additional degrees of freedom, the {\em
localized phasons} or {\em elementary flips}, which consist of local
rearrangements of tiles (groups of 4 tiles in dimension 3). The
activation of these degrees of freedom gives rise to a large amount of
accessible configurations which are responsible for a macroscopic
configurational entropy, the calculation of which is in itself a
difficult topic that the present paper does not address directly.

Among the many problems that remain unsolved in this field, results on
flip-dynamics are scarce in three
dimensions~\cite{Tang90,Shaw91,Jaric94,Gahler95,Ebinger98,Linde01} and
are either purely numerical or based upon an approximate Langevin
approach. And yet it is a crucial issue in quasicrystal science where
elementary flips are believed to play an important role because they
are a new source of atomic mobility.  They could bring their own
contribution to self-diffusion~\cite{kalugin93} in quasicrystalline
alloys and they are involved in some specific mechanical properties,
such as plasticity related to dislocation mobility~\cite{wollgarten93}
(see subsection~\ref{physical_diff} for a more
detailed discussion).

The present paper addresses two issues related to flip dynamics:
connectivity of tiling sets {\em via} elementary flips and
self-diffusion (of vertices) in random tilings when flips are
activated.

The question of the connectivity of tiling sets {\em via} elementary
flips still resists investigation in spite of the apparent simplicity
of its formulation: {\em is it possible to reach any tiling from any
other one by a sequence of elementary flips?} Even if proving that
tiling sets are connected {\em via} elementary flips is only a first
step towards the full characterization of flip dynamics, it is a
challenging question that must imperatively be addressed before
tackling more complex issues such as ergodicity, self-diffusion,
calculation of ergodic times~\cite{PRLbibi,Linde01} or dislocation
mobility. This connectivity issue is also crucial in the context of
Monte Carlo simulations on tilings: it is a fundamental ingredient if
one hopes to sample correctly their configuration spaces.

So far, connectivity has only been conjectured by Las Vergnas about 25
years ago in the context of ``oriented matroid
theory''~\cite{LasVergnas80}. It remains an open problem in pure
mathematics~(\cite{Reiner99}, Question 1.3). In two dimensions,
connectivity can be established~\cite{kenyon93,elnitsky97}, but the
proofs are very specific to dimension 2 and cannot be adapted to
dimensions 3 and higher. In reference~\cite{octo01}, a new proof of
the connectivity in dimension 2 was proposed and the reason why this
proof could not be easily extended to higher dimensions was clearly
identified: there appear ``cycles'' (defined below) in the generalized
partition method which locally suppress a fraction of flip degrees of
freedom.  It is this point of view that we shall adopt in the present
paper: we shall demonstrate that the obstacles to the generalization
of the latter proof can be rigorously bypassed in many cases: in
codimensions 1 and 2; in most cases in codimensions 3 and 4. We say
``most cases'' because a new difficulty arises when one studies
three-dimensional rhombic tilings: for a given codimension, all edge
orientations are not equivalent. There are 4 non-equivalent edge
orientations in codimension 3 and 11 ones in codimension 4. The paper
discusses this non-trivial issue in great detail. There is a minority
of edge orientations where we are not able to prove connectivity. Note
that when all edge orientations are not combinatorially equivalent,
the corresponding polyhedral fixed boundaries are not topologically
equivalent either. This point will also be discussed in detail in the
paper.

Beyond connectivity, the previously mentioned cycles are likely to
affect ergodicity. Since they suppress some flip degrees of freedom,
they are susceptible to slow down the dynamics and to be responsible
for entropic barriers which could for instance prevent a tile from
finding its average equilibrium position in the tiling. We have chosen
to study self-diffusion of vertices to explore such possible effects
because of its physical interest. Self-diffusion has previously been
studied in icosahedral tilings~\cite{Jaric94,Gahler95,Ebinger98} but
the effect of cycles themselves has never been investigated. We shall
see that there are no significant differences in the diffusive behavior
between tilings where cycles exist and those where we are able to
prove that they cannot exist, nor between tilings where connectivity
can be proven and those where it remains open. No sub-diffusive
regimes at long time are observed whatever the tilings under
consideration. Cycles do not seem to slow down the dynamics, at least
as far as self-diffusion is concerned.

The paper is organized as follows: In section 2, we describe the
``generalized partition'' method used throughout the paper to code the
tilings and we explain how ``cycles'' emerge in this formalism.  In
the following section 3, we set the basics of flip dynamics, and we
discuss the possible influence of cycles on the dynamics by
flips. Section 4 discusses the question of non-equivalent edge
orientations and related fixed boundaries. In section 5, we give our
main theorem that states that cycles cannot exist in favorable
conditions, which enables us to prove connectivity by flips in a large
variety of cases. When cycles exist, we also study how abundant they
are and we derive a simple mean-field argument to account for our
observations. In addition, we make a brief incursion into order
theory: we prove that the tiling sets have a structure of ``graded
poset''. In section 6, we discuss how our results on fixed-boundary
tilings can be transposed to the more physical free-boundary ones.
Finally, section 7 is devoted to a numerical study of the
diffusion of vertices. The last section 8 contains conclusive remarks
and open questions.

\section{Tilings, generalized partitions and cycles}

In this section, we present the concept of generalized partition
used in the paper to code and manipulate rhombus tilings. This
technique was introduced in~\cite{elser84,mosseri93,mosseri93B},
developed in~\cite{destain97,octo01} and mathematically formalized
in~\cite{bailey97}. The end of the
section is devoted to the definition of cycles. We provide
a commented example.

\subsection{Generalities}
\label{gene}

\begin{figure}[h!] 
\begin{center}
\resizebox{8cm}{!}{\includegraphics{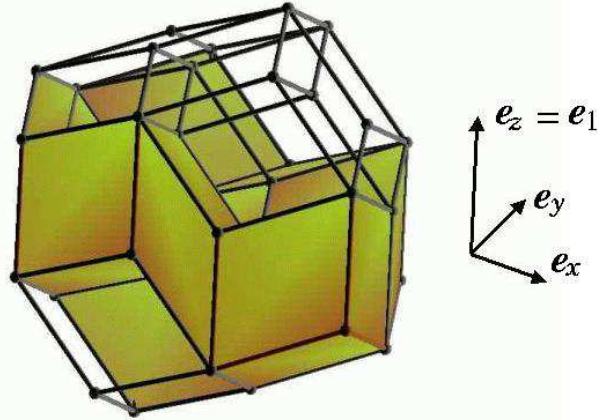}}
\end{center}
\caption[]{Unitary $6 \ra 3$ tiling with icosahedral symmetry. Shaded
tiles form the de Bruijn surface attached to the edge orientation
$\evect{1}$ equal to $\evect{z}$ in this figure. This de Bruijn
surface can be seen as a (mono-valued) function from $(xOy)=\Rb^2$ to
$(Oz)=\Rb$.  It is topologically equivalent to a $5 \ra 2$ tiling with
decagonal boundary.  }
\label{f_surf3D}
\end{figure}

The rhombus tilings considered in this paper (see an example in
figure~\ref{f_surf3D}), whatever their dimension, have $D$ possible
edge orientations (belonging to $\Rb^d$), denoted by $\evect{a}$,
$a=1,\ldots,D$. They are inherited from the $D$ directions of the
cubic lattice of $\Rb^D$ during the ``cut-and-project'' process. Each
possible edge has the orientation and norm of one of the vectors
$\evect{a}$. The signs of these vectors $\evect{a}$ are irrelevant and
can be arbitrarily chosen. A rhombic tile is defined by $d$ of these
edge orientations. To avoid flat tiles, the family of orientation
vectors is supposed to be {\em non-degenerate}: any $d$ of them form a
basis of $\Rb^d$.

A dual representation of rhombus tilings was introduced by de
Bruijn~\cite{DeBruijn1,DeBruijn2}. It consists in seeing the tiling as
a grid of lines (see figure~\ref{f_octopartition}, right). A line in a
tiling is a succession of adjacent tiles sharing an edge (in dimension
2) or a face (in dimension 3) with a given orientation. It is always
possible to extend these lines through the whole tiling up to a
boundary tile. These lines are called ``de Bruijn lines''.  In
dimension $3$, one can also define de Bruijn surfaces which can be
represented by adjacent tiles sharing an edge with a given orientation
$\evect{a}$ (see figure~\ref{f_surf3D}). They will play an
important role below. It exists one family of
surfaces, denoted by $F_a$, for each orientation $\evect{a}$
of edges. There are $p_a$ surfaces in the family $F_a$. De Bruijn
surfaces of the same family do not intersect.  A tiling with $D$
orientations of edges on a $d$-dimensional space will be called a $D
\ra d$ tiling, and $D-d$ defines its codimension. In dimension $d$,
the de Bruijn surfaces are replaced by $(d-1)$-dimensional
hyper-surfaces.

In this paper, we will mainly be interested in the case $d = 3$, and
we shall use $2$-dimensional tiling examples to facilitate the
comprehension. In dimension $3$, a rhombic tile is given by the
intersection of three surfaces of different families, and so all the
different types of tiles are given by all the possible intersections
of $3$ surfaces of different families; there are ${D \choose 3}$
different species of tiles. We can prolong any de Bruijn surface of
family $F_a$ beyond the tiling boundaries and up to infinity, by a
surface perpendicular to the direction $\evect{a}$ far from the
tiling. If the tiling corresponds to a complete grid, which means that
any three surfaces of different families have a non-empty
intersection, it will have fixed boundary conditions. More precisely,
this boundary will be a zonotope~\cite{coxeter}, that is to say the
shadow of the hypercube of sides $(p_1,p_2,\dots,p_D)$ in the
$D$-dimensional space onto the $d$-dimensional space, where we recall
that the $p_a$ are the numbers of hyper-surfaces in family $F_a$. We
shall describe more precisely this kind of boundaries in
section~\ref{boundaries} and we shall discuss their relationship
with free boundaries in section~\ref{free_fixed}. 

Here we make an important remark about de Bruijn surfaces: a
$(d-1)$-dimensional de Bruijn surface $S$ in a $D \ra d$ tiling $t_d$
is topologically equivalent to a $D-1 \ra d-1$ tiling. For example, a
de Bruijn surface in a 3-dimensional tiling can also be seen as a
2-dimensional tiling. Indeed, let us consider the trace on $S$ of the
$d$-dimensional de Bruijn grid dual of $t_d$: it is a grid made of
$D-1$ families of $(d-2)$-dimensional surfaces. This grid is complete
because the original grid is. It is the dual of a $D-1 \ra d-1$ tiling
with fixed zonotopal boundaries. In figure~\ref{f_surf3D}, a de Bruijn
surface is represented. The equivalent 2-dimensional rhombus tiling is
obtained by looking at the surface from the direction $\evect{z}$.

A tiling will be called ``unitary'' if it contains one de Bruijn
surface per family ($p_a=1$ for all $a$). It will be
called ``diagonal'' if it contains the same number of 
de Bruijn surfaces in each family ($p_a=p$ for all $a$).

\subsection{Generalized partitions}

Here we introduce the notion of generalized partitions which plays a 
central role in the paper. We also explain how an edge orientation
$\evect{a}$ orients the faces of a tiling. This notion will 
be fundamental in the proof of connectivity.

The idea of the generalized partitions method is to build iteratively
the tiling, by reconstructing the dual grid. In dimension $3$,
beginning with a complete grid made by only three families of
surfaces, which is unique and represents a $3 \ra 3$ periodic tiling
made by one type of tiles, we have to describe where to place the de
Bruijn surfaces of the fourth family $F_4$ relatively to the existing
intersections of $3$ surfaces. In terms of tilings, it means that we
have to describe where to place the fourth surfaces of tiles on the
existing $3 \ra 3$ tiling. And iteratively, to build a $D+1 \ra 3$
tiling, we have to describe where to place the family $F_{D+1}$ of
surfaces on a previously obtained $D \ra 3$ tiling. In order to obtain
a tiling by this process, we have to impose some constraints on the
way we place the next family of surfaces at each step. The surfaces of
one family cannot intersect. No more than $3$ surfaces can cross at
the same point, otherwise the tile at this point cannot be properly
defined. Furthermore, de Bruijn surfaces are
directed~\cite{destain97}. Indeed, by construction, they always cross
edges with a given orientation, and can be seen as mono-valued
functions $\Rb^{d-1} \ra \Rb$ defined on the hyper-plane perpendicular
to their orientation vector $\evect{a}$ (see figure~\ref{f_surf3D}).

\begin{figure}[h!] 
\begin{center}
\resizebox{8cm}{!}{\includegraphics{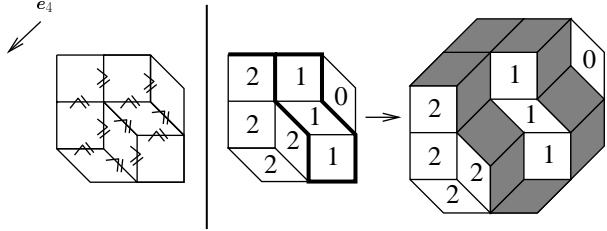}}
\end{center}
\caption[]{Example of generation of a $4 \ra 2$ tiling by the
generalized partition method. Left: Order relation between the tiles
of a base $3 \ra 2$ tiling $\tilde{t}$. Right: one solution of this
generalized partition problem codes a $4 \ra 2$ tiling. The tiles of
the de Bruijn family $F_4$ are shaded. The remaining white ones belong to
the base tiling $\tilde{t}$. The domains $D_0$, $D_1$ and $D_2$ appear
as the tiles bearing parts equal to 0, 1 and 2
respectively. They are separated by the two de Bruijn surfaces of the
family $F_4$. From reference~\cite{octo01}.}
\label{f_octopartition}
\end{figure}

Let us now introduce (see figure~\ref{f_octopartition}) how one can
define a partial order relation between the tiles of a tiling in order
to satisfy these constraints. Since the $p_a$ de Bruijn surfaces of a
family $F_a$ do not intersect, they divide the space $\Rb^d$ in
$p_a+1$ disjoint domains.  Furthermore, since de Bruijn surfaces are
globally oriented, we can index these domains from 0 to $p_a$ such
that, following the direction given by $\evect{a}$ we go through all
these domains in an increasing order. We denote these domains by
$D_0,\ldots,D_{p_a}$ and the surfaces of $F_a$ by
$S_1,\ldots,S_{p_a}$.  The de Bruijn surface $S_k$ lies between the
domains $D_{k-1}$ and $D_k$. In other words, if we consider a tiling
$t$, and if we particularize the de Bruijn surfaces of the family
$F_a$ (see figures~\ref{f_surf3D} and \ref{f_octopartition}), tiles not
belonging to the surfaces of $F_a$ are distributed between the
different domains.

Now we contract (or delete) the tiles of $F_a$ from the $D+1 \ra d$
tiling $t$ by setting the length of $\evect{a}$ to 0, thus obtaining a
$D \ra d$ tiling $\tilde{t}$.  Two adjacent tiles of $\tilde{t}$, with
one above the other along the direction $\evect{a}$, are either on the
same domain $D_k$ or separated by one (or several) de Bruijn surface
of $F_a$, the tile atop being in the higher domain. This allows to
define an order relation, $\leq_a$, relatively to $F_a$, between any
two adjacent tiles $u$ and $v$ in $\tilde{t}$: $u \leq_a v$ means that
$u$ is below $v$ along $\evect{a}$ and so that $u$ is either in the
same domain as $v$ or in a lower domain.

We can recover this order relation between any adjacent tiles by {\em
orienting} all the faces\footnote{In this paper we call face of a tile
a $(d-1)$-dimensional polyhedron generated by $d-1$ orientation
vectors.} of a tiling $\tilde{t}$ by $\evect{a}$: given two adjacent
tiles $u$ and $v$ of $\tilde{t}$, $u \leq_a v$ if when one goes from
$u$ to $v$, the face between $u$ and $v$ is crossed in the positive
direction (see figure~\ref{f_cycle_example}). We say that the vector
$\evect{a}$ orients the faces of $\tilde{t}$.

Now remind that our aim is to code the position of the de Bruijn
surfaces of $F_{D+1}$ on a $D \ra 3$ tiling $\tilde{t}$. One way to do
that, is to associate an integer $X_u$, $0 \leq X_u \leq
p_{D+1}$, called a {\em part}, to each
tile $u$ of the tiling $\tilde{t}$: $X_u$ is equal to the index $k$ of
the domain $D_k$ the tile $u$ belongs to.  For example, one tile with
a part equal to $3$ is on the third domain, so between the second and
the third de Bruijn surface of the family $F_{D+1}$; A tile with a
part equal to zero is below the first surface.  These parts have to
respect the partial relation order $\leq_{D+1}$, which for convenience
we will simply denote by $\leq$ in the following of this paper. To
generate all the possible $D+1 \ra 3$ tilings from a $D \ra 3$ one, we
have to find all the possibilities of filling the tiles of the latter
tiling by parts from $0$ to $p_{D+1}$ respecting the partial order
between the tiles. This type of problem is called a generalized
partition problem of height $p_{D+1}$ on the $D \ra 3$ tiling. A
solution of this problem is called a (generalized) partition.  In the
figure~\ref{f_octopartition}, one can see an example of a $4 \ra 2$
tiling coded by a generalized partition on a $3 \ra 2$ tiling. The
underlying tiling $\tilde{t}$ is called the {\em base} tiling of the
generalized partition problem.

To sum up, there is a one-to-one correspondence between $D+1 \ra 3$
tilings and pairs composed of a base $D \ra 3$ tiling together with a
generalized partition on this base tiling. This one-to-one coding of
zonotopal tilings is described in a more formal way in references
\cite{bailey97,octo01}. It can be iterated by induction on $D$ to code
$D \ra 3$ tilings, starting from the simplest case of partitions on $3
\ra 3$ tilings. The latter can be seen as 3-dimensional rectangular
arrays and the corresponding partition problems are usually called 
solid partition problems~\cite{destain97}.

To close this section, let us remark that when one codes $D+1 \ra 3$
tilings by the generalized partition technique, the order in which the
de Bruijn families of surfaces are successively added to the tilings
is arbitrary. For sake of convenience, one is for example free to
choose the $D$ first edge orientations defining the base tilings among
the $D+1$ possible ones.  Of course, the set of base tilings depends on
this choice.

\subsection{Definition and examples of cycles}

\begin{figure}[h!] 
\resizebox{\textwidth}{!}{
\begin{tabular}{|c|c|c|}
\hline
\resizebox{!}{5cm}{\includegraphics{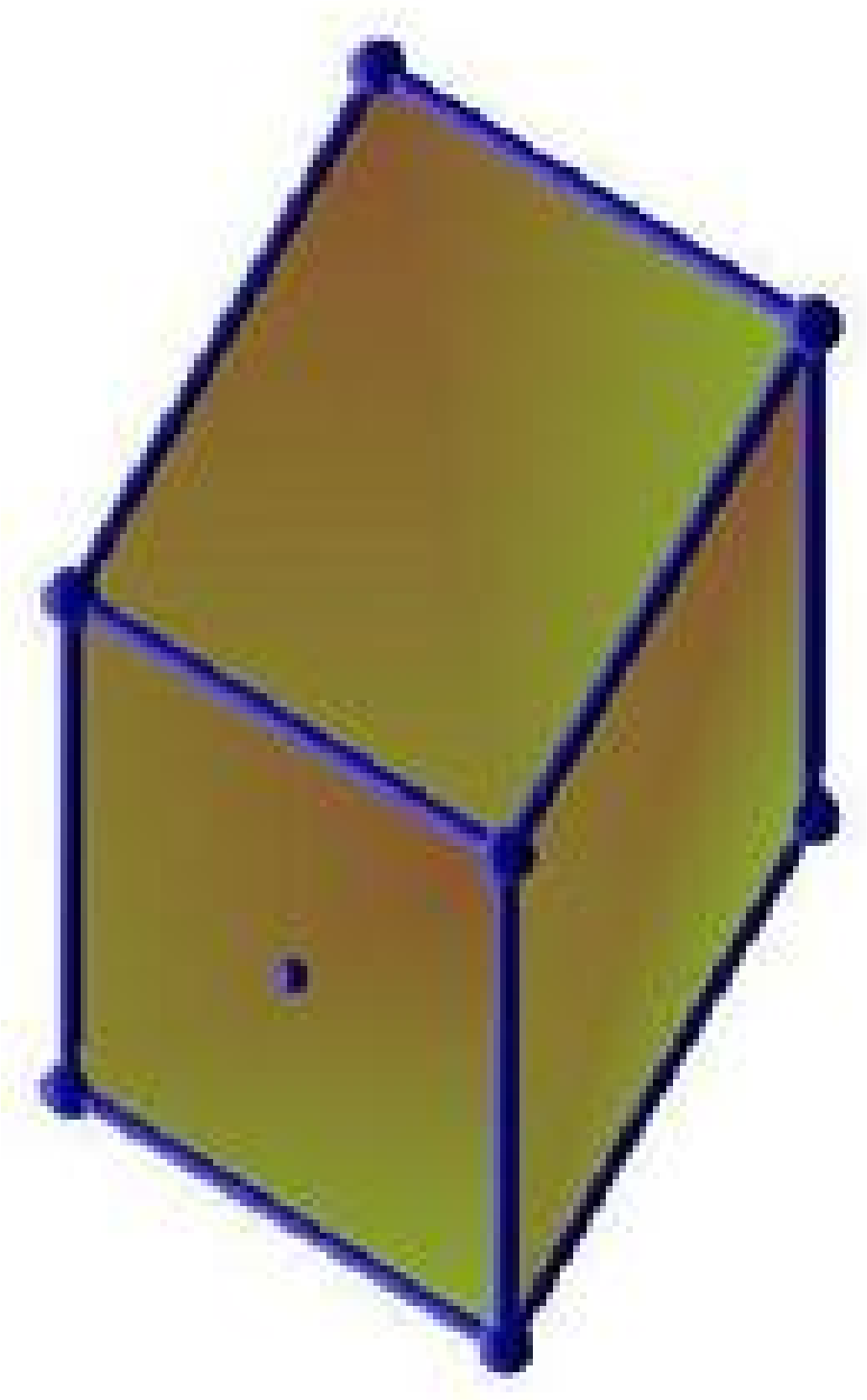}} &
\resizebox{!}{5cm}{\includegraphics{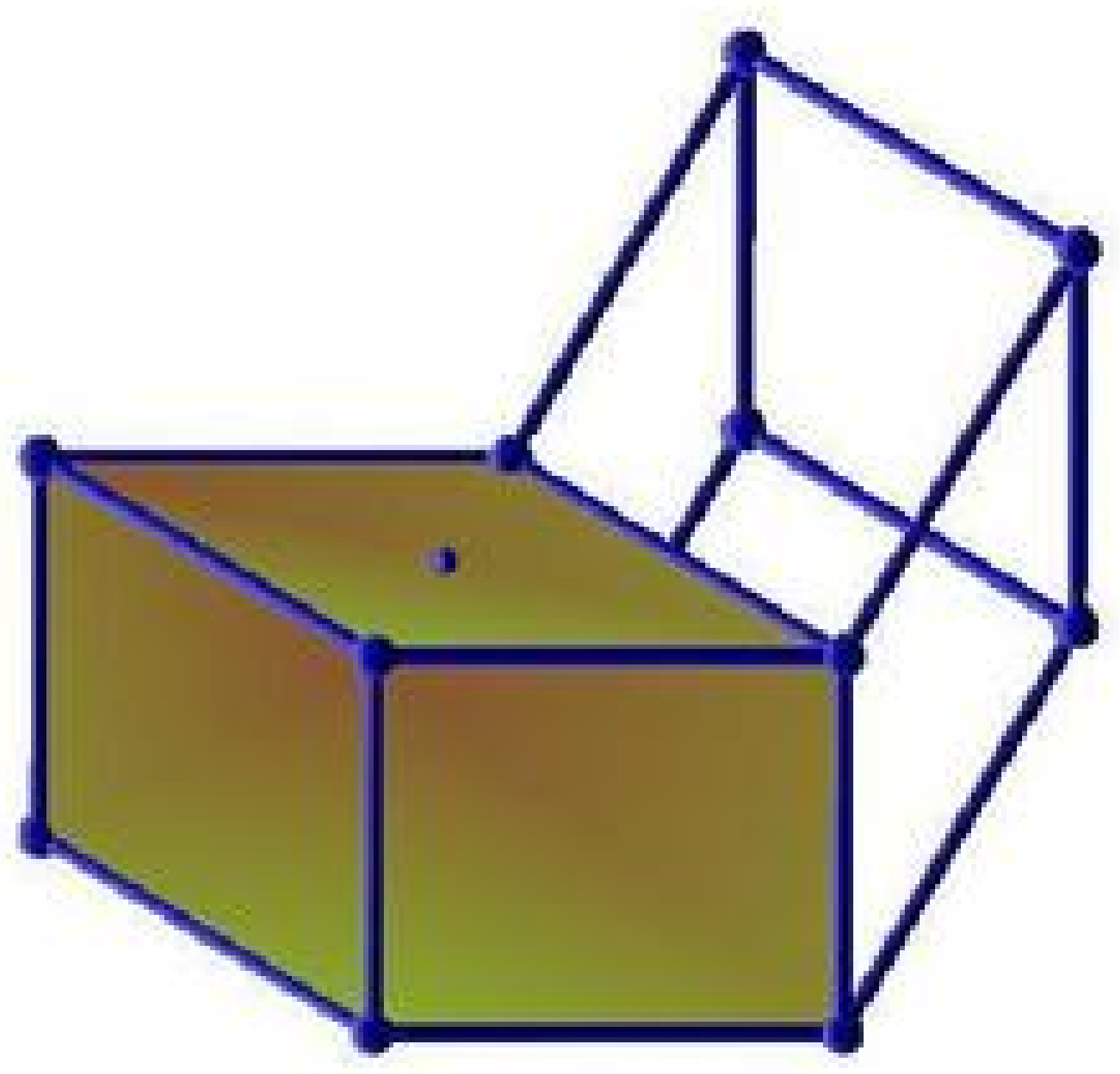}} &
\resizebox{!}{5cm}{\includegraphics{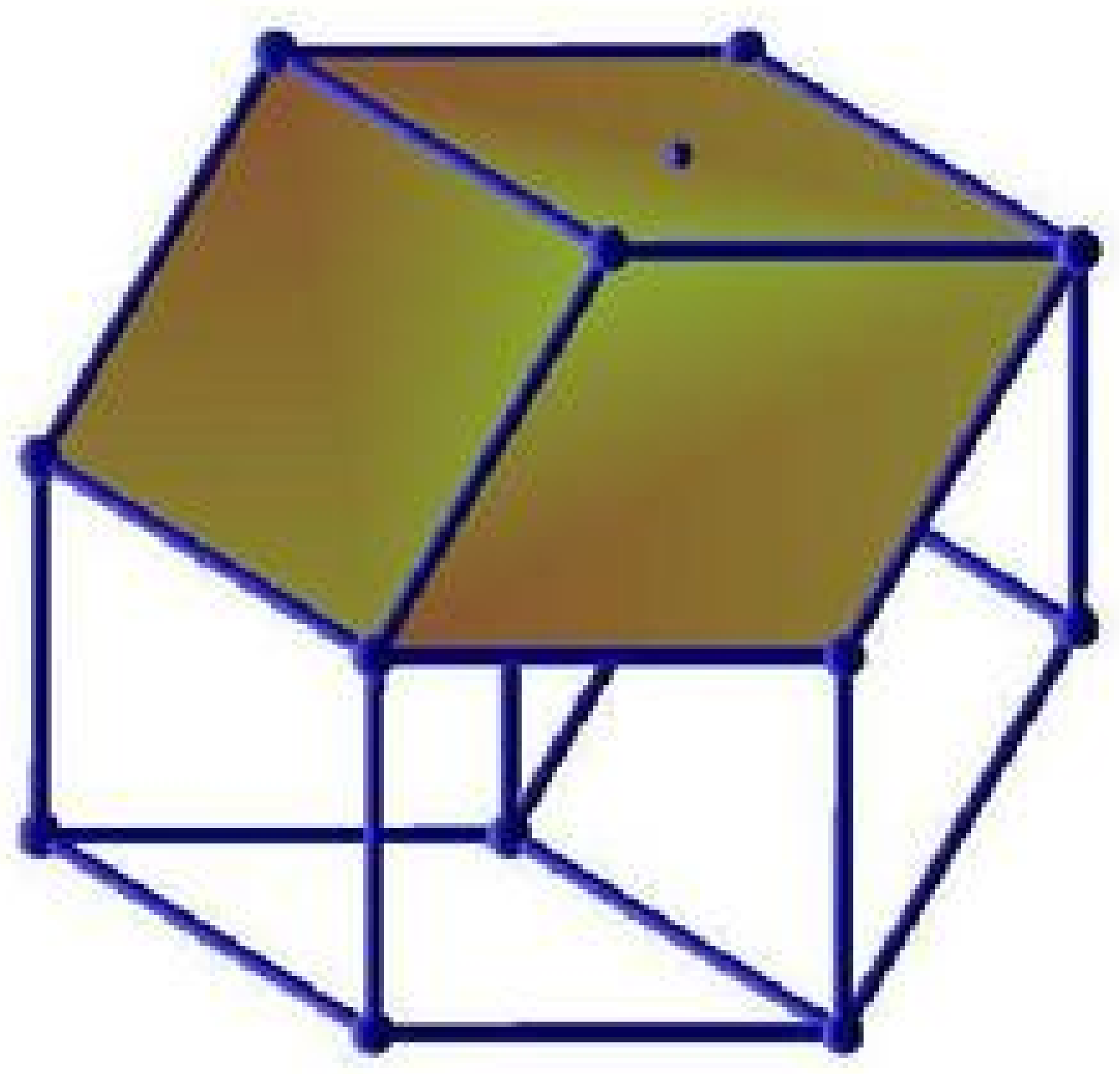}} \\
\hline
\resizebox{!}{5cm}{\includegraphics{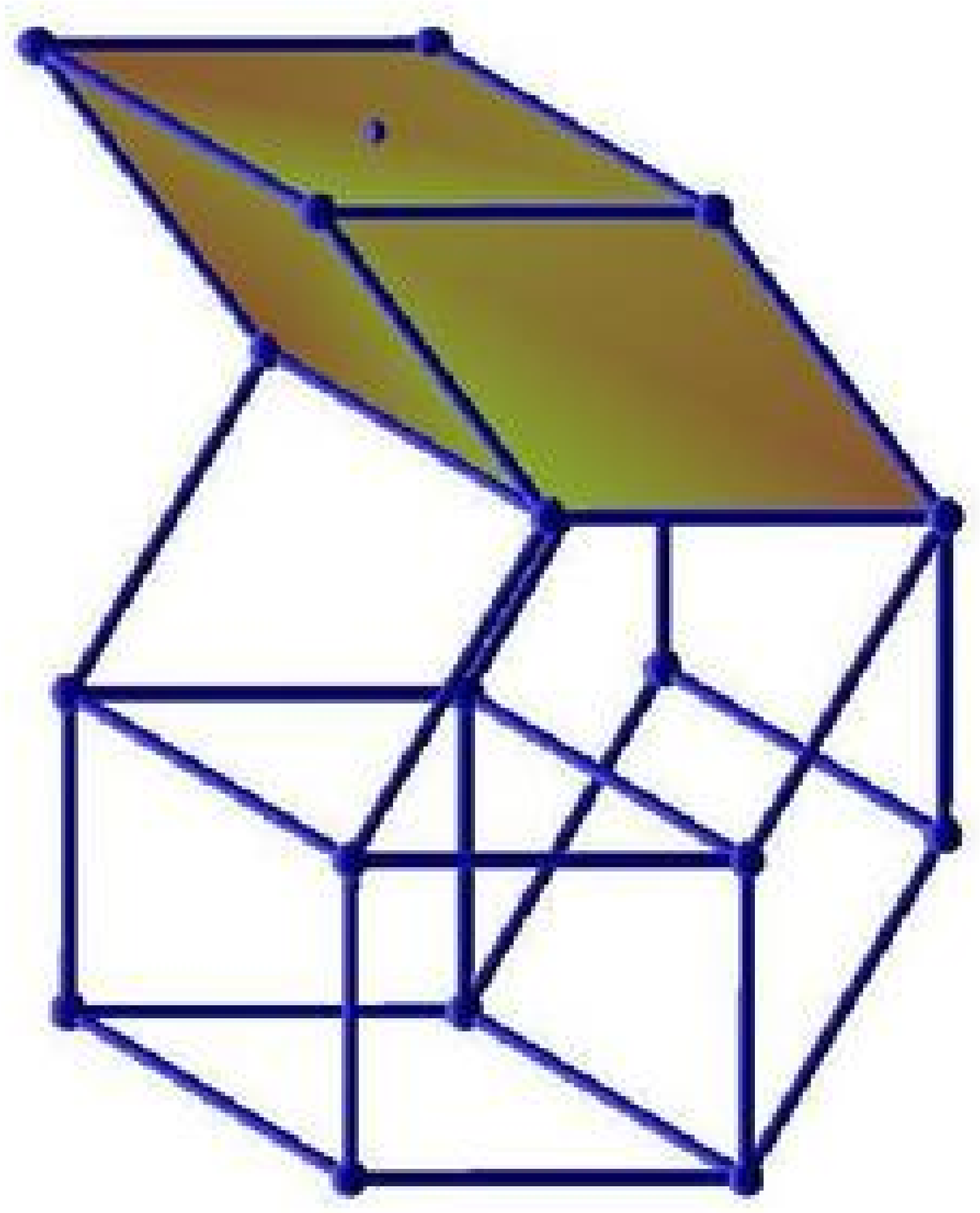}} &
\resizebox{!}{5cm}{\includegraphics{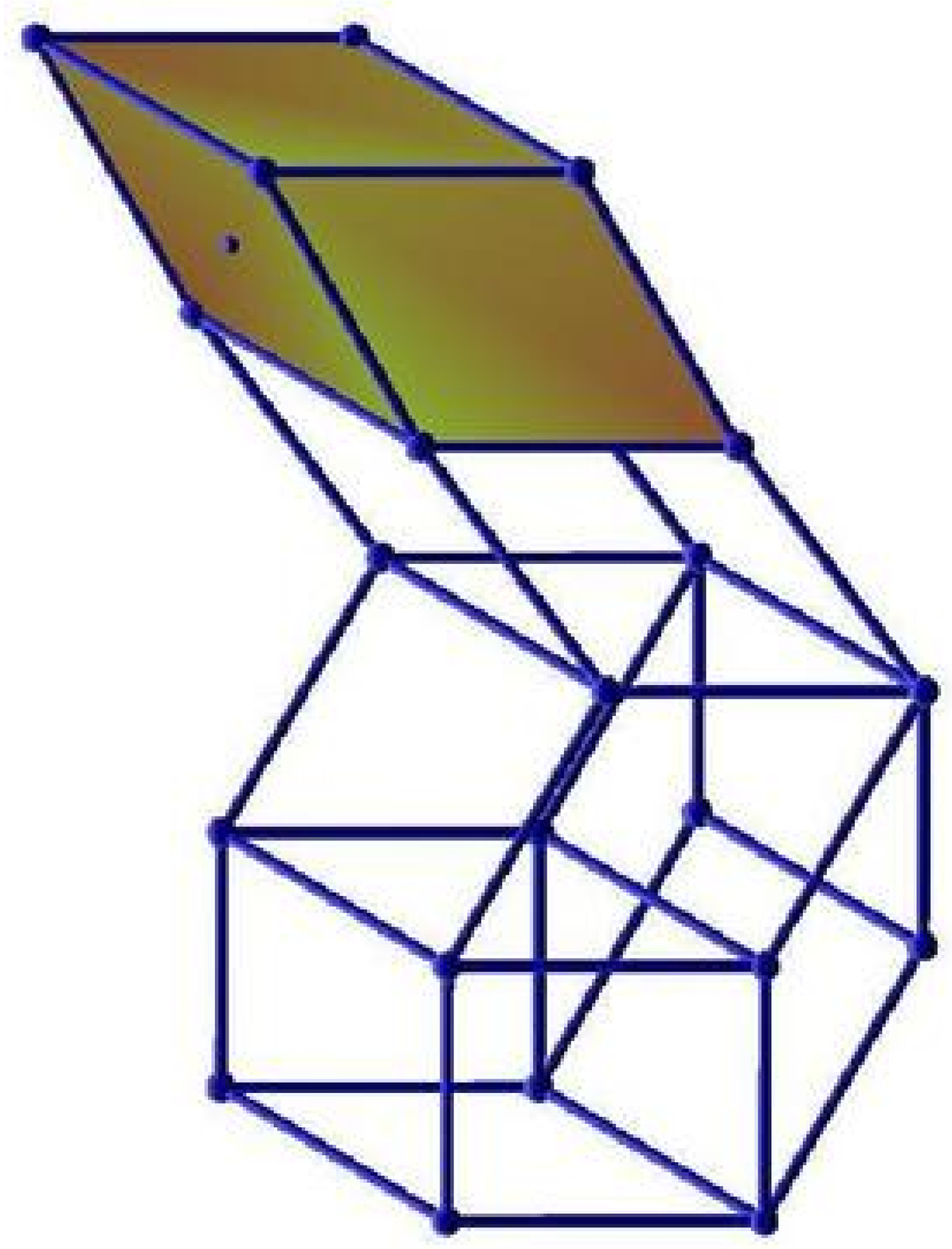}} &
\resizebox{!}{5cm}{\includegraphics{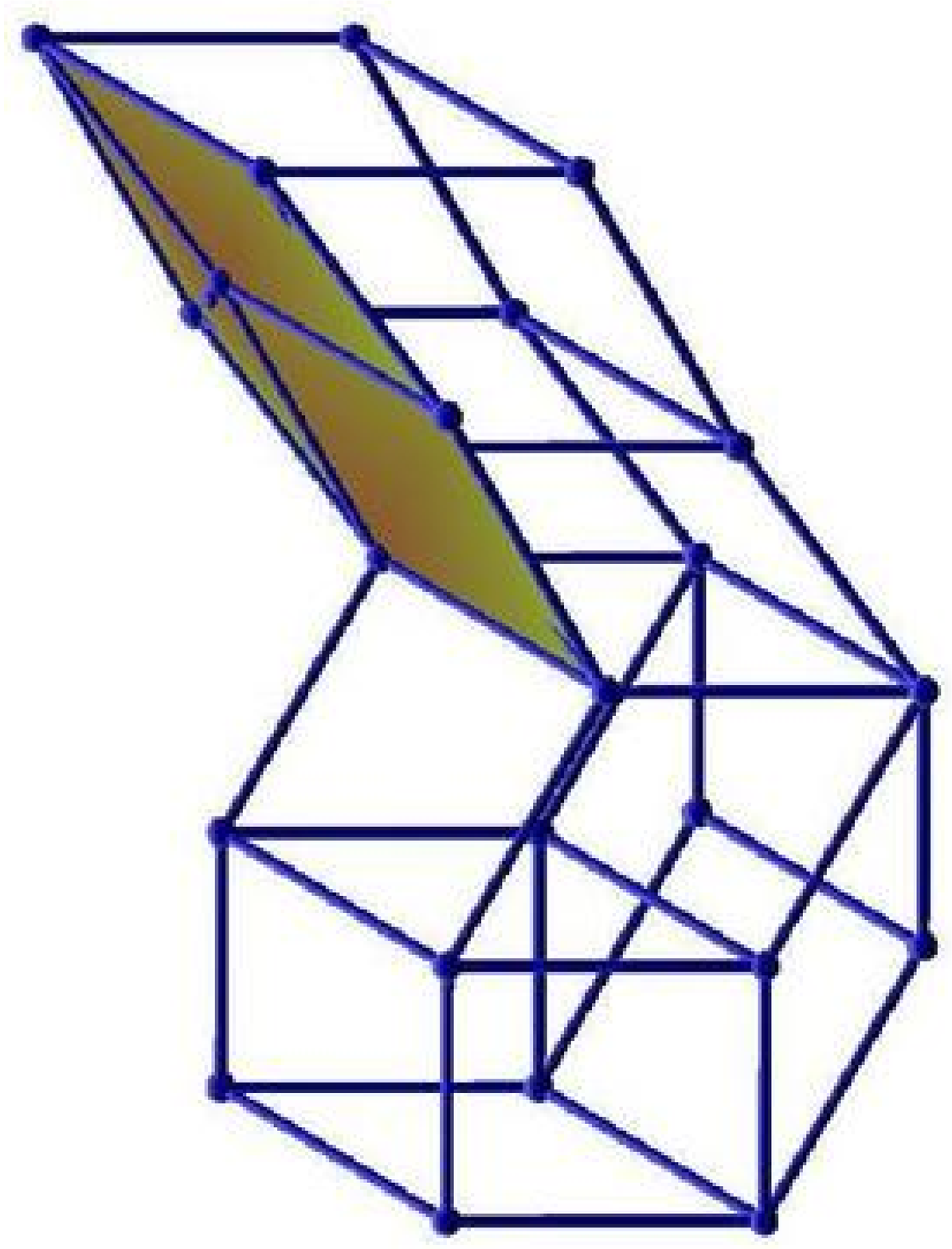}} \\
\hline
\resizebox{!}{5cm}{\includegraphics{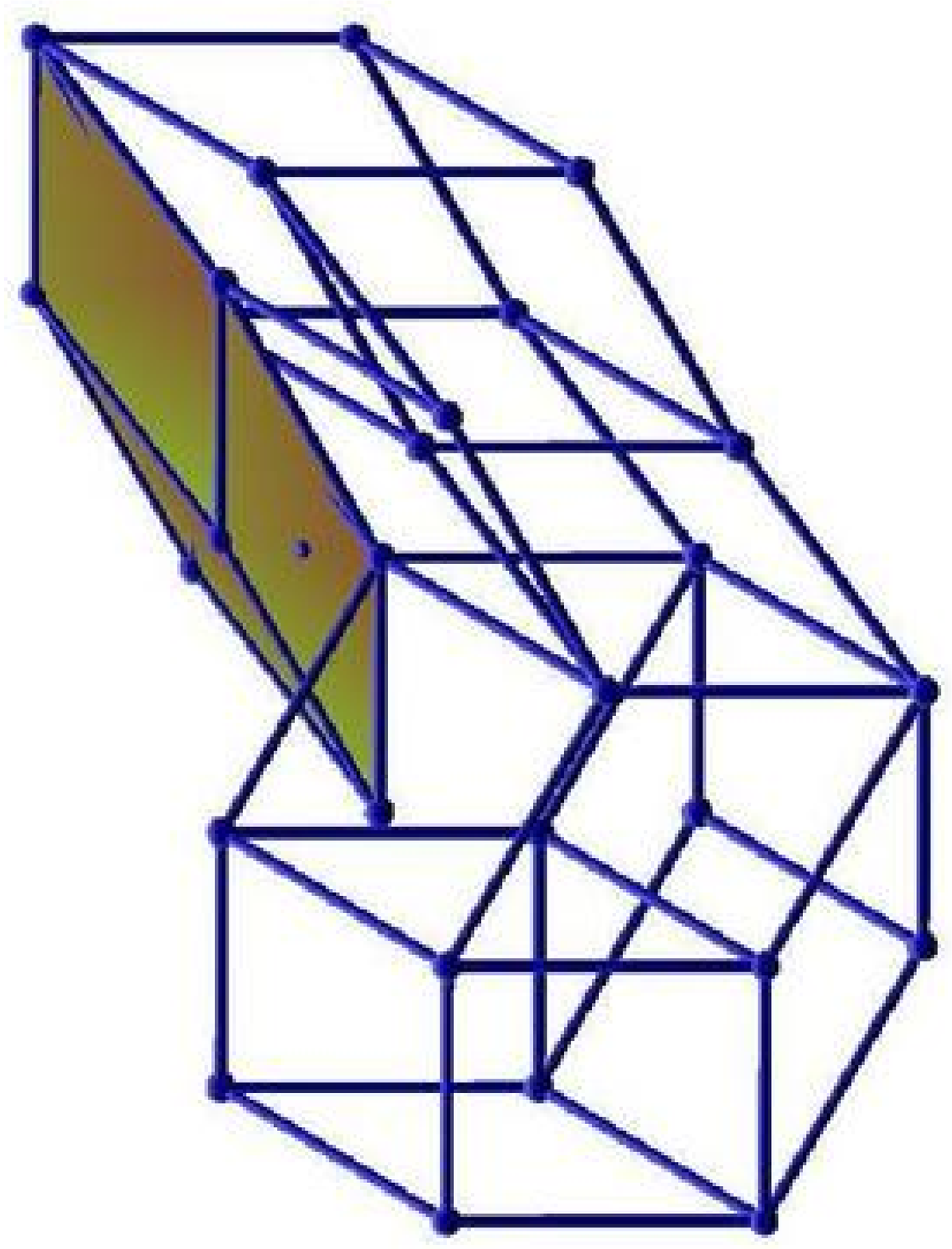}} &
\resizebox{!}{5cm}{\includegraphics{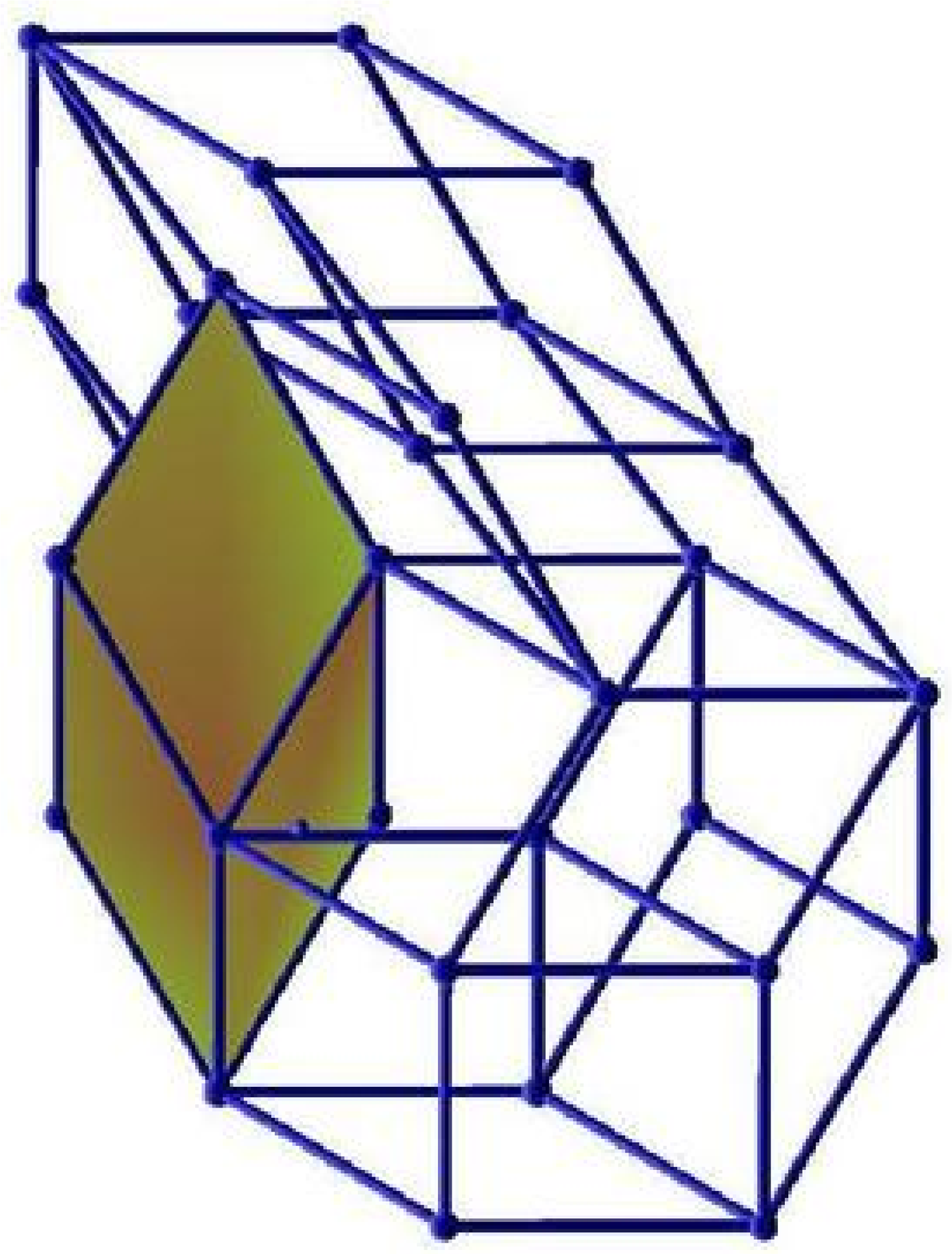}} &
\resizebox{!}{5cm}{\includegraphics{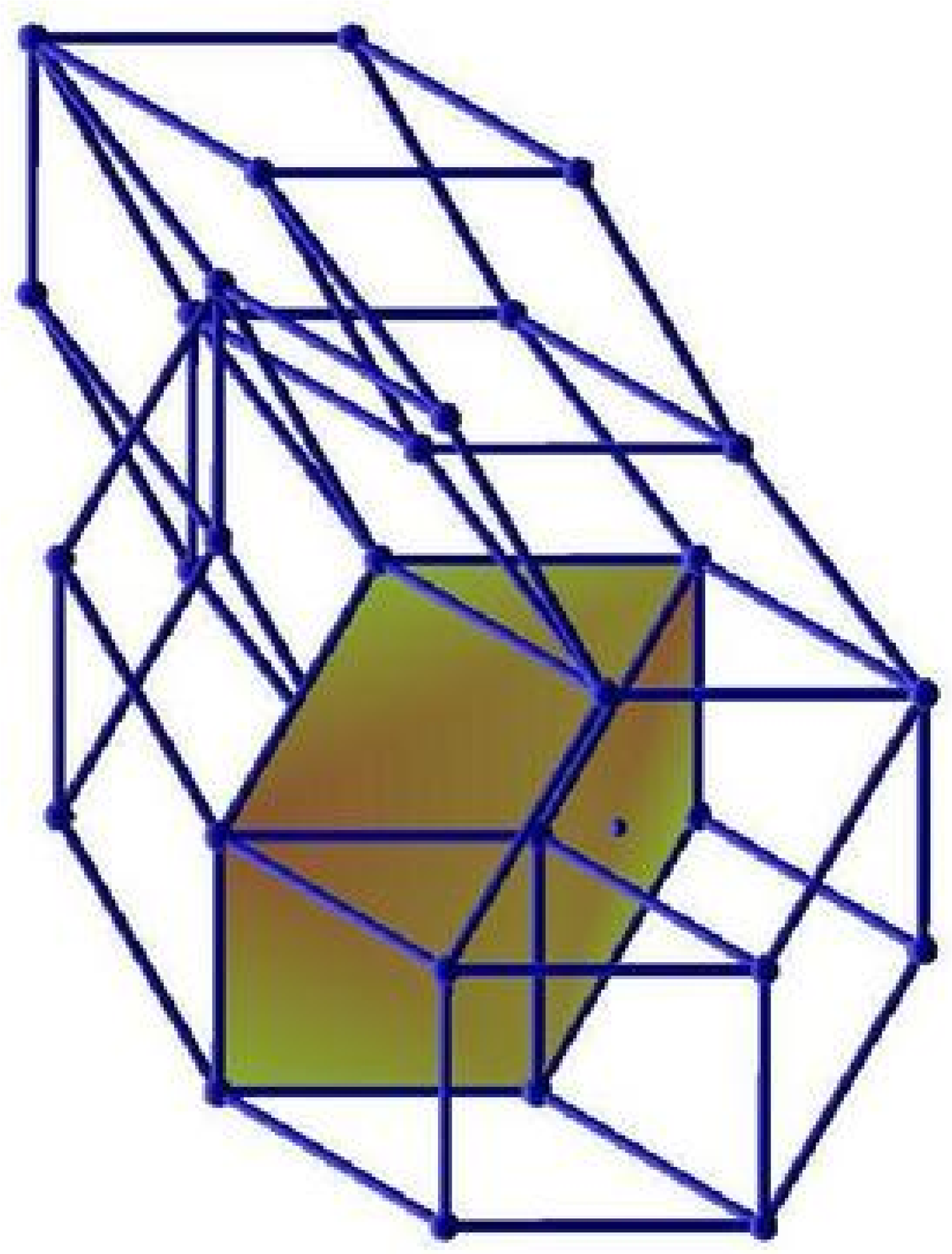}} \\
\hline
\end{tabular}
}
\caption[]{Example of 9-tile cycle in a base unitary $6 \ra 3$ tiling
with one de Bruijn surface in each family. We have only drawn the 9
tiles belonging to the cycle whereas the tiling contains 20
tiles. Such cycles already exist in $5 \ra 3$ tilings but the examples
we know contain more tiles than the present one. The tiles are added
to the cycle one by one. The edge orientation $\evect{7}$ which
orients the tiling is perpendicular to the picture plane and points
upwards.  As a consequence, given two adjacent tiles $u$ and $v$
separated by a tiling face, $u$ is just above $v$ with respect to
the order relation between tiles if $u$ is above $v$ in the
figure, in other words if $u$ hides partially $v$ (the face of
$u$ that will be covered by $v$ is marked by a dot).  Each tile of
the cycle is above the preceding one and the last tile is below the
first one, which loops the cycle. Note that {\em this} cycle can be
broken by flipping the 4 bottom tiles that form a rhombic
dodecahedron.}
\label{f_cycle_example}
\end{figure}

We can now define what we call a {\em cycle} on a base
tiling~\cite{octo01}. A cycle is a succession of pairwise adjacent
tiles, $u_1, u_2, \dots, u_n, u_1$, such that: $X_{u_1} \leq X_{u_2}
\leq \dots \leq X_{u_n} \leq X_{u_1}$ with respect to the previous
order relation on tiles, as it is illustrated in
figure~\ref{f_cycle_example}. That means that the parts of the tiles
inside the cycle have to be equal to a unique part $X_0$ and have a
collective behavior. In particular, their values cannot but change
simultaneously. Such cycles are known to exist on specific {\em ad
hoc} $6 \ra 3$ examples. The first one can be found in reference
\cite{Bjorner93} (example 10.4.1) and the second one
in~\cite{Sturmfels93} (example 3.5), in the context of ``oriented
matroid theory''. Figure~\ref{f_cycle_example} provides another
example. Our analysis below shows that cycles already exist in $5 \ra 3$
tilings, but not in unitary ones. A base tiling with cycles is said to
be {\em cyclic}, and conversely a tiling without cycles is {\em
acyclic}.

Geometrically speaking, a cycle is a sequence of tiles making a loop
such that each tile is placed above the preceding one relatively to
the orientation prescribed by $\evect{D+1}$. In the example of
figure~\ref{f_cycle_example} this vector is placed perpendicularly to
the picture plane. This cycle can be seen analogously to a loop of
coins, each one placed above the preceding one.  A visible consequence
is that we cannot place a de Bruijn surface of $F_{D+1}$ between the
tiles of the cycle, as well as we cannot place a horizontal sheet of
paper between the coins, splitting the loop between coins above the
sheet and below the sheet. A de Bruijn surface of $F_{D+1}$ is either
completely above or completely below a cycle. Which is equivalent to
say that the tiles must bear equal parts.

One of our purposes below is to analyze the occurrence of cycles in a
more systematic way. In addition, in the following of this article, we
will discuss the possible influence of these cycles on the
configuration spaces of tilings and on flip dynamics.  Let us
emphasize that this problem is specific to dimensions $3$ and above,
since there cannot exist cycles in dimension
$2$~\cite{Bjorner93,Sturmfels93}.


\section{Basics of flip dynamics} \label{s_cycle}

Before discussing in further details the consequences of these
cycles, we now describe flips in rhombus tilings, and what they
become in the generalized partition viewpoint.

\begin{figure}[h!] 
\begin{center}
\begin{minipage}{4cm}
\includegraphics[width=4cm]{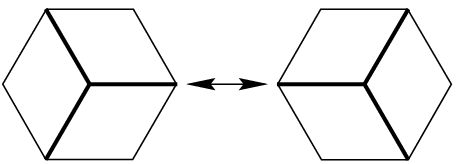} \\
\end{minipage}
\hspace{0.5cm}
\begin{minipage}{8cm}
\includegraphics[width=8cm]{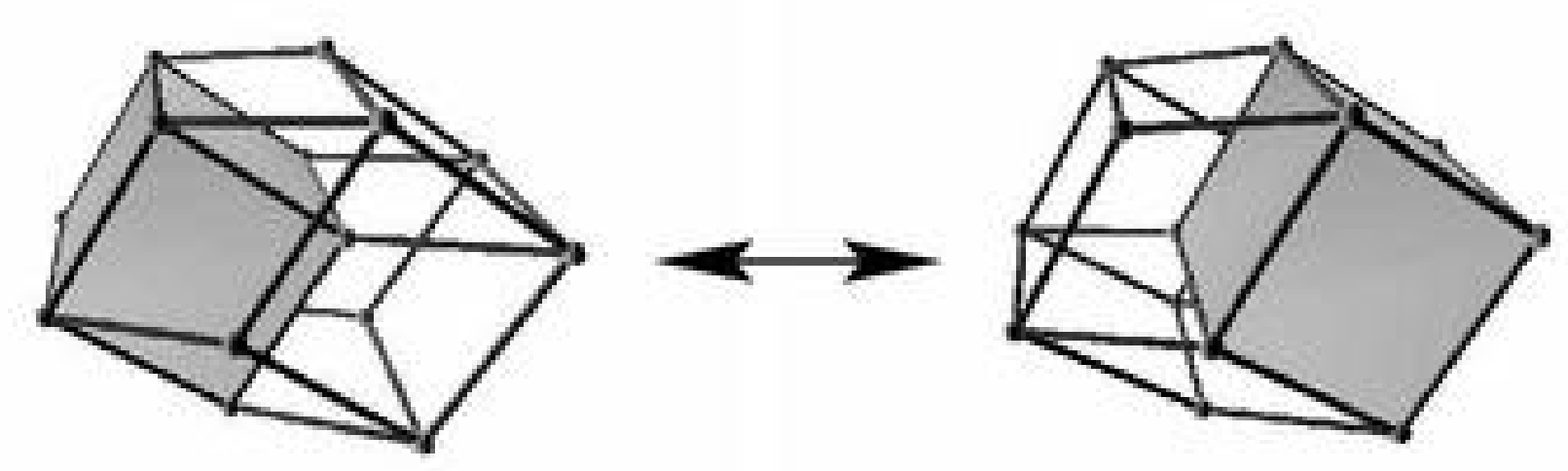} \\
\end{minipage}
\end{center}
\vspace{-1cm}
\caption[]{A 2-dimensional flip inside a (not necessarily regular)
hexagon involves 3 tiles, and a 3-dimensional flip inside a rhombic
dodecahedron involves 4 tiles.}
\label{f_flip}
\end{figure}

\begin{figure}[h!]
\vspace{-2cm}
\begin{center} 
\resizebox{7cm}{!}{\includegraphics{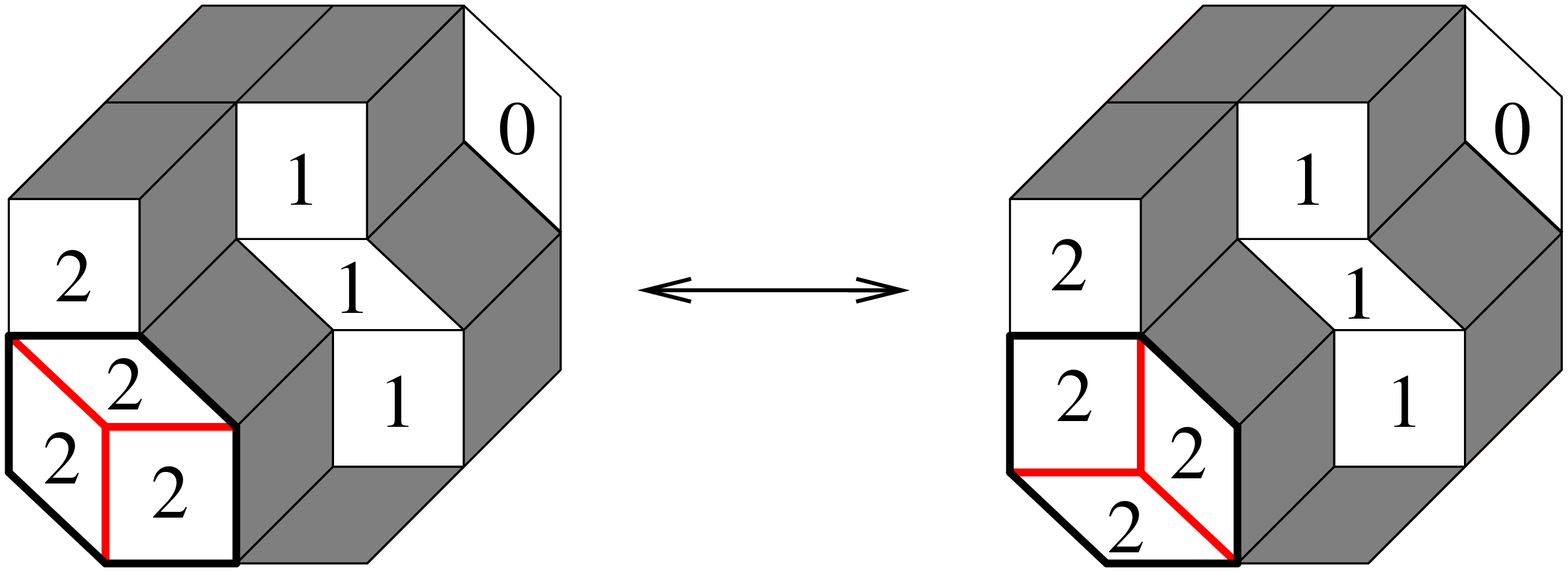}} \\
\vspace{-0.5cm}
\resizebox{7cm}{!}{\includegraphics{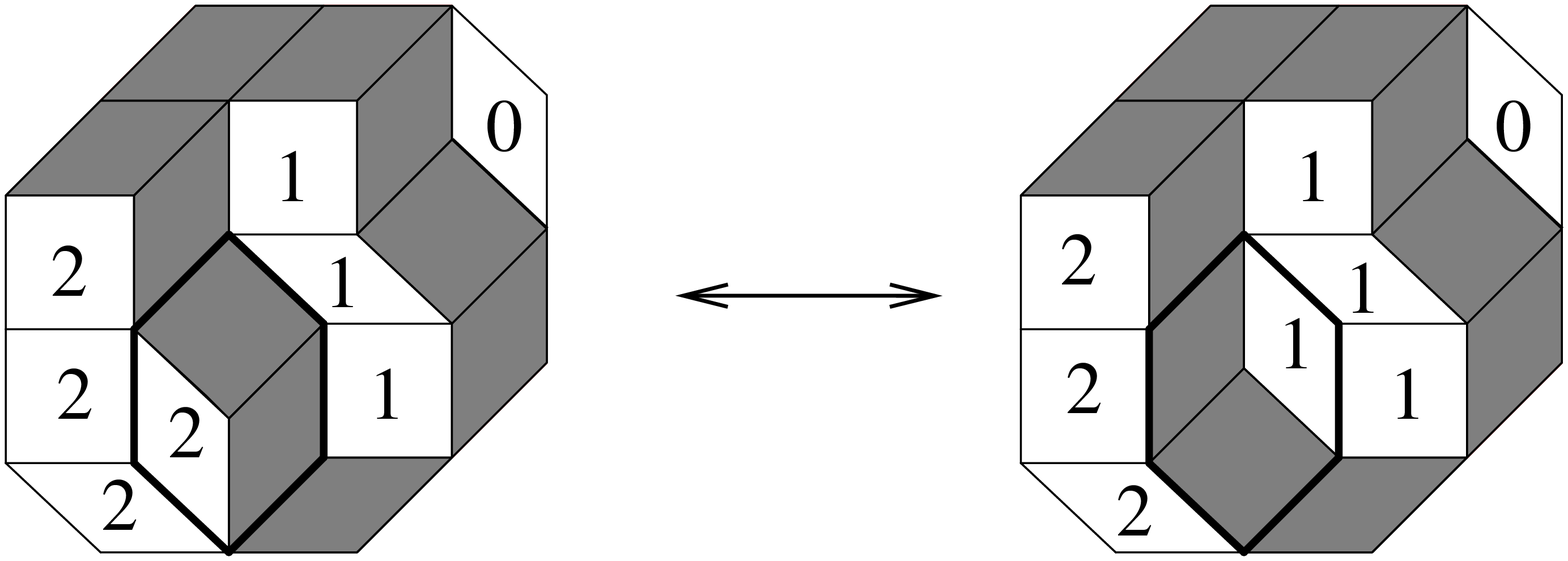}}
\end{center}
\caption[]{The two types of flips on the $4 \ra 2$ tiling of
figure~\ref{f_octopartition}. Upper panel: type-I flip involving 3
tiles none of them in the family $F_4$. It only affects the base
tiling but not the parts it bears. Lower panel: type-II flip involving
2 tiles of the family $F_4$ and one single tile $u$ of the base
tiling. It changes the part $X_u$ borne by $u$ by $\pm 1$.}
\label{f_flip_partition}
\end{figure}

One can define in rhombus tilings local degrees of freedom which are
called elementary flips or localized phasons.  In dimension $d$, a
flip consists of a local rearrangement of $d+1$ tiles in a small
zonotope inside the tiling. In dimension $2$ it is a rearrangement of
$3$ tiles inside a hexagon, and in dimension $3$, of $4$ tiles inside
a rhombic dodecahedron, see figures~\ref{f_flip}
and~\ref{f_flip_partition}. In dimension $2$ the configuration space
of tilings is proven to be connected {\it via} these elementary flips
\cite{kenyon93,elnitsky97,octo01}. Which means that we can go from any tiling to any other
one by a finite sequence of flips.  In dimension $3$, it is an open
question that we shall address in this paper.

Flips allow to define a Monte Carlo Markovian dynamics on tiling sets
as follows~\cite{Tang90,Strandburg90}: pick up a tiling vertex at
random with uniform probability. If this vertex is flippable (it is
surrounded by $d+1$ tiles in dimension $d$), then flip it.  This
Markovian process converges towards the uniform distribution on the
set of tilings provided the configuration space is connected by
flips. Note that temperature can be introduced in this point of view
to take into account possible interactions between tiles; the
transition rates must be adapted consequently. In this paper, we focus
on the infinite temperature limit, where all the configurations have
equal equilibrium probability and where all rates of allowed
transitions are equal.

This Markovian dynamics has been mainly studied in dimension 2.  It
has been demonstrated that it is rapid in codimensions
1~\cite{Randall98} and 2~\cite{PRLbibi}. This means that the typical
times to reach equilibrium are polynomial in the system size. The same
kind of result has also been established numerically in the $4 \ra 3$
case~\cite{Linde01}. In addition, there exist studies concerning
diffusion in random tilings evolving {\em via} this Markovian
dynamics. This last point will be discussed in great detail in
section~\ref{diffusion}.

Now, let us see how these flips are seen in the generalized partition
point of view.  On a generalized partition problem on a $D \ra d$
tiling, which codes a $D+1 \ra d$ tiling, flips can be classified into
two types (see figure \ref{f_flip_partition}), following
reference~\cite{octo01}.

Type-I flips involve only tiles of the base $D \ra d$ tiling and no
tile of the de Bruijn family $F_{D+1}$. As a consequence, the $d+1$
tiles bear equal parts, and these flips only change the base tiling
without modifying the parts of the tiles. Type-II flips involve tiles
belonging to the de Bruijn family $F_{D+1}$. More precisely, they
involve $d$ tiles having an edge oriented by $\evect{D+1}$, locally
representing a surface of the family $F_{D+1}$, and one tile $u$ of
the base tiling. Such a flip consists in modifying the position of the
latter surface with respect to the tile $u$. So it changes the part
$X_u$ borne by the tile $u$ by $\pm 1$. 

\begin{figure}[h!] 
\begin{center}
\resizebox{!}{8cm}{\includegraphics{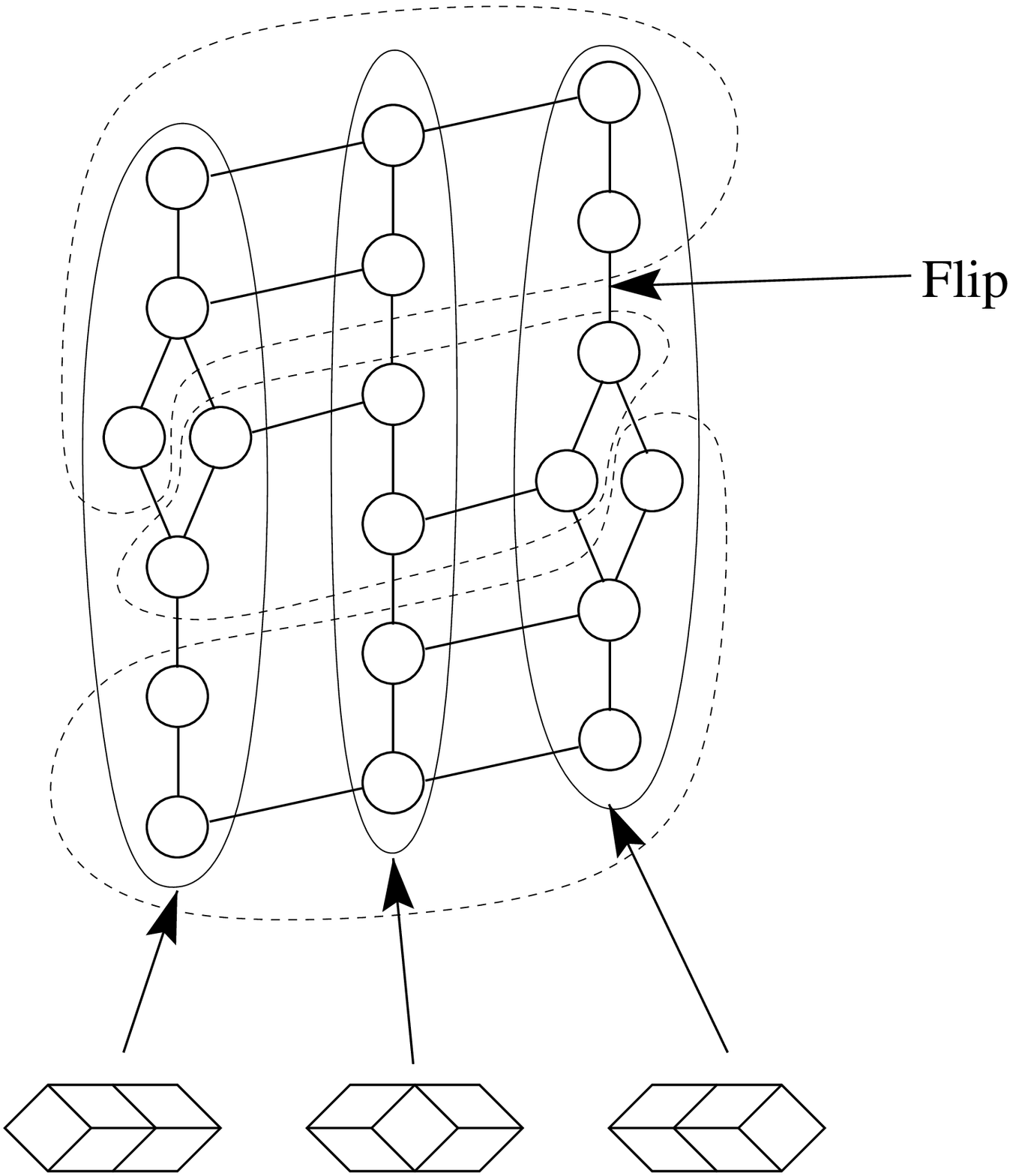}}
\end{center}
\caption[]{Configuration space of $4 \ra 2$ tilings of an octagon of
sides (2,1,1,1). There are 20 tilings represented by vertices. Tilings
are linked by an edge if they differ by a single flip. In the
generalized partition formalism, the 20 tilings can be distributed
among 3 fibers according to their base tiling.  The 3 possible $3 \ra
2$ base tilings are represented in the figure. One can see
type-I (inter-fiber) flips and type-II (intra-fiber) flips. If one
particularizes a different edge orientations from the forth family of
lines, the fibration is different. There are 4 different fibrations
corresponding to the 4 edges $\evect{a}$. We have represented two of
them (fibers drawn in full and dotted lines, respectively).

}
\label{f_space_fiber}
\end{figure}

One can now build a schematic picture of the configuration space of
tilings~\cite{octo01} (see figure~\ref{f_space_fiber}). Considering
$D+1 \ra d$ tilings, one can split up the configuration space into
disjoint {\it fibers}. A fiber contains all the tilings generated by
the same generalized partition problem, that is to say which have the
same base tiling. Type-II flips keep the base tiling, and therefore the
fiber, unchanged, whereas type-I flips change the base tiling and
therefore the fiber. The set of all fibers is called a {\em fibration}.
For a given configuration space, there are $D+1$ different fibrations
corresponding to the choice of $\evect{D+1}$ among the $D+1$
edge orientations.

Using this picture, one immediately gets the following result: {\em if
fibers are all connected and if the base is connected itself, then the
configuration space is connected in its turn}. Indeed, the
connectivity of fibers allows one to put all the tiles of the $D \ra
d$ tiling to the same part value (for example 0), thus releasing all
type-I flips on the base tiling and allowing to go to any fiber.

Now it is established in~\ref{connex_fib} that a fiber corresponding
to an acyclic base tiling is connected, because it is possible 
to change the parts on tiles one by one. Therefore {\em if all base
tilings are acyclic and if the base is connected, then the
configuration space is connected in its turn}. Since there cannot
exist cycles in dimension 2, sets of $D \ra 2$ tilings are always
connected~\cite{octo01}. We shall follow this route in the following
to prove connectivity in a wide variety of cases.

What happens when {\em there are} cycles? A cycle in a tiling is a
sequence of pairwise adjacent tiles that are geometrically constrained
to bear the same part in the generalized partition problem. In term of
flips, since all parts of the cycle are forced to be equal, the tiles
of a cycle cannot participate to a type-II flip which would change the
part of a {\em single} tile to a value different from that of the
whole cycle.

In other words, a de Bruijn surface of $F_{D+1}$ cannot pass through a
cycle, because it is constrained to be completely above or completely
below the cycle.  In order to allow the surface to pass the cycle, one
must break beforehand the cycle by type-I flips, if it is
possible. The tiling is locally {\em jammed}.

\begin{figure}[h!] 
\begin{center}
\resizebox{!}{7cm}{\includegraphics{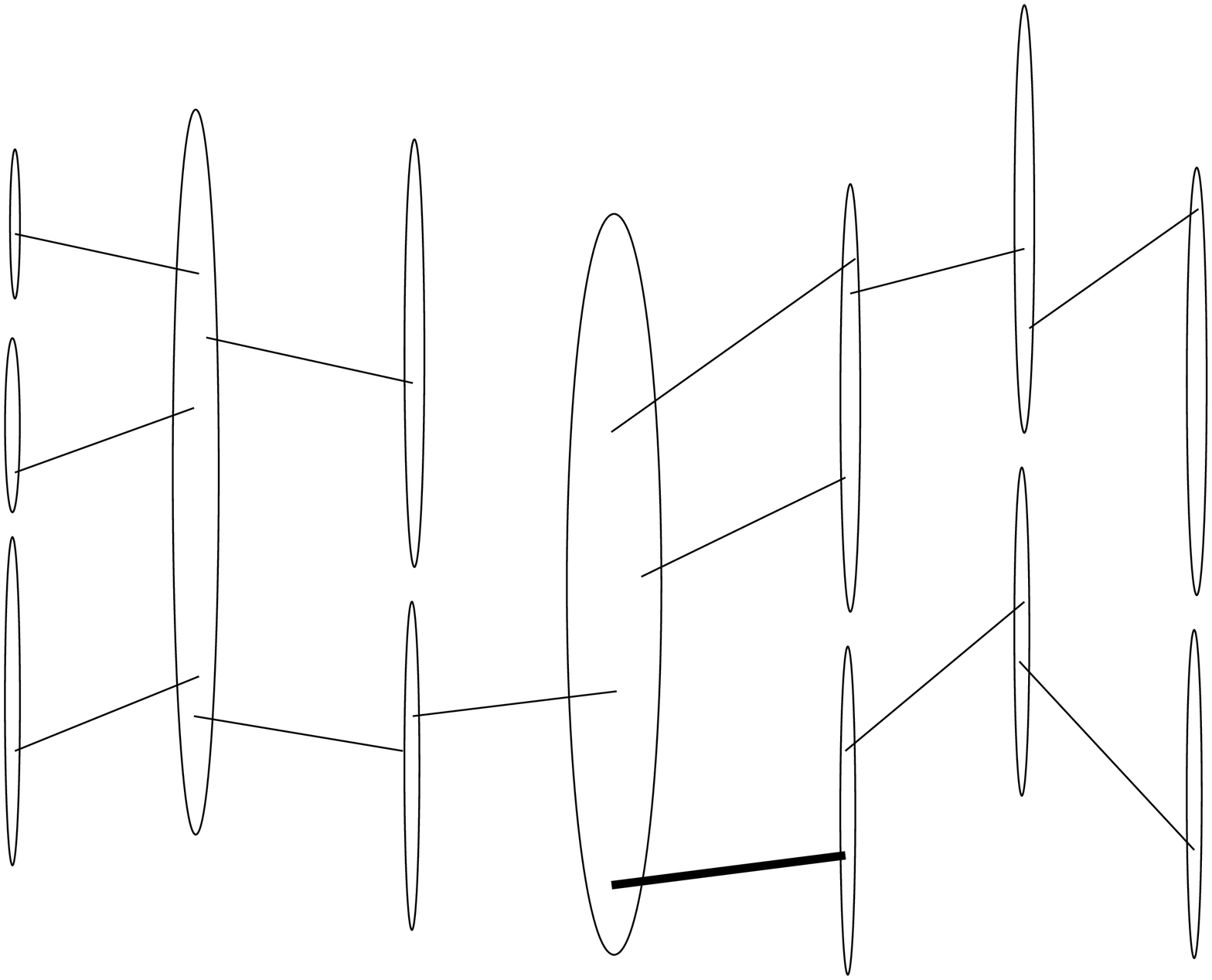}}
\end{center}
\caption[]{Schematic picture of a configuration space where fibers can
be disconnected because of the possible occurrence of cycles. Some
fibers are connected if they are associated with acyclic base tilings,
some are not. This configuration space if more intricate than the one
of figure~\ref{f_space_fiber} and its overall connectivity is not
acquired any longer. In particular in this figure, if the flip
represented by the thick line is suppressed, the connectivity fails,
the space is split into two components.}
\label{f_fiber_space_cycle}
\end{figure}

A direct consequence is that a fiber based on a tiling with cycles
cannot be connected anymore by single flips. We must modify our
schematic picture of the configuration space: there are connected
fibers based on tilings without cycle, as well as disconnected ones
based on tilings with cycles. The connectivity is not obvious anymore
(see figure \ref{f_fiber_space_cycle}).

Let us anticipate on the following to emphasize that these cycles
should have an influence not only on the connectivity but also on the
Markovian flip dynamics. Indeed they forbid some type-II flips, they
reduce locally the degrees of freedom related to those flips because
of jammed clusters of tiles. They are susceptible to slow down the
dynamics, in the sense of an increase of ergodic times. In the
schematic picture of figure~\ref{f_fiber_space_cycle}, one can see
that the cycles make the configuration space more intricate. One can
imagine that they could create inhomogeneities in the distribution of
flip paths through the phase space resulting in entropic
barriers. More precisely, in \cite{PRLbibi}, the demonstration of
short ergodic times in dimension $2$ was based on short ergodic times
inside the fibers, which is not possible anymore in presence of
cycles. Therefore, one can wonder whether the features of the dynamics
are modified by those cycles.  It will be the purpose of
section~\ref{diffusion}.

\section{Edge orientations, line arrangements and boundaries}
\label{boundaries}

Before tackling the questions of connectivity and vertex diffusion,
this section clarifies the question of non-equivalent edge
orientations in 3-dimensional tiling problems and their relation with
zonotopal boundaries. Indeed, it will appear in the following that the
existence (or not) of cycles in base tilings is closely related to the
choice of edge orientations.  In particular, we shall be able to prove
connectivity by flips for a large majority of edge orientations, but
the proof will fail in some minority cases. The classification of edge
orientations will be related to a classification of line arrangements
in the projective plane $\Pb\Rb^2$.

\subsection{Edge orientations and equivalence relation}

Random tiling model studies concern the way of arranging simple
geometrical structures (the tiles) in the space or in the plane. One
possible interest is the contribution to the entropy of such possible
configurations.  Here we are interested in the way the system can go
from one of those configurations to another. The geometry of the tiles
does not concern directly those features but rather the
symmetries. Indeed, one can imagine to start with a tiling and begin
to vary slightly one vector $\evect{a}$. It will deform
globally the tiling, but not its topological structure in terms of
relative positions of the tiles. More precisely, vectors $\evect{a}$ can be
rotated or elongated provided a modified vector does not cross a plane
made to by two other ones, which means there exist no flat tiles and
no tiles are overlapping. 

Given two families of $D$ edges, denoted by
$f=(\evect{1},\ldots,\evect{D})$ and
$f'=(\evect{1}',\ldots,\evect{D}')$, we define them as {\em
equivalent}~\cite{ziegler_b95} if one can transform $f$ into $f'$ by
the composition of the three following transformations: (i)
permutation of the indices; (ii) sign reversals; (iii) continuous
deformation of the vectors $\evect{a}$ without creating degenerate
configurations of three vectors:
$\det(\evect{a_1},\evect{a_2},\evect{a_3}) \neq 0$ for all
$(a_1,a_2,a_3)$. In dimension 2, all families of $D$ edges are
equivalent. When the families $f$ and $f'$ are equivalent, we say that
they define equivalent sets of rhombic tiles.

\subsection{Line arrangements}

To distinguish and enumerate the non-equivalent families of edge
orientations, we map families of edges on line arrangements in the
projective plane $\Pb\Rb^2$ (see~\cite{ziegler_b95} for more
details). We proceed as follows.  The set of edges are represented by
a family of $D$ vectors $(\evect{1},\ldots,\evect{D})$. Recall that
the signs of those vectors are irrelevant. Those families of vector
arrangements are in bijection with arrangements of planes $(\HC_{1},
\HC_{2}, \dots,\HC_{D})$, such that for all $a$, $\evect{a}$ is
orthogonal to $\HC_{a}$ and each $\HC_{a}$ contains the origin. For
one given arrangement, since all the planes pass through a common
point ({\em i.e.} the origin), one can get all the information on it
by its trace on the projective plane $\Pb\Rb^2$ (which is conveniently
represented by an affine plane that does not contain the origin. Then
this trace is made of lines $(\LC_1,\LC_2,\ldots,\LC_D)$ defined as the
intersections of the projective plane and the $\HC_{a}$). So, one can
differentiate vector arrangements by differentiating the line
arrangements in the projective plane.

\begin{figure}[h!] 
\resizebox{\textwidth}{!}{\includegraphics{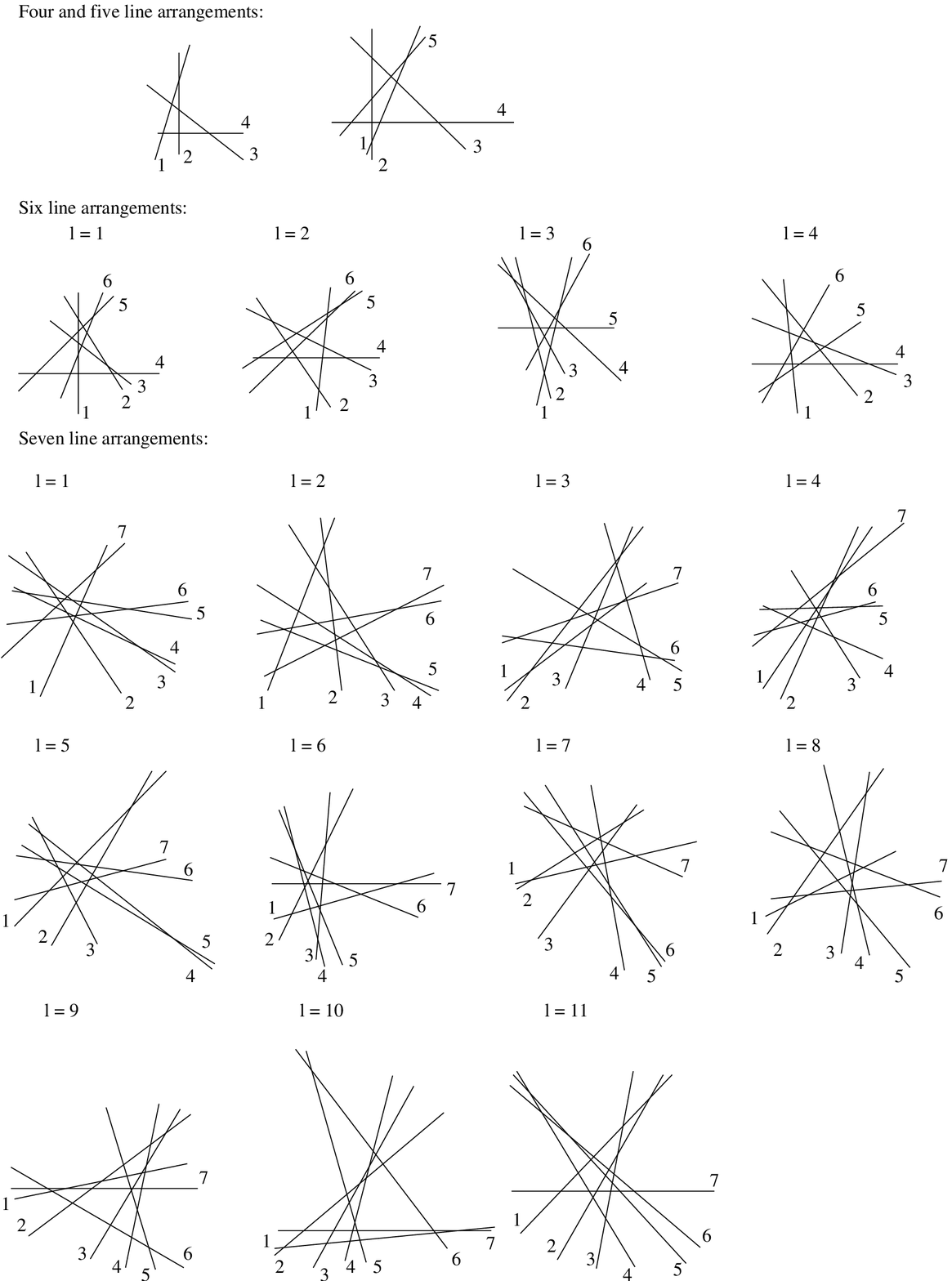}}
\caption[]{Equivalence classes of line arrangements in the projective
plane~\cite{grunbaum_b67}. These line arrangements are put in the text
in bijection with edge orientations and three-dimensional zonotopal
boundaries. There are respectively 1, 1, 4 and 11 arrangements of 4,
5, 6 and 7 lines. The first arrangement of six lines corresponds to
the icosahedral symmetry.}
\label{f_linesarrangement}
\end{figure}

We precise now in the line arrangement point of view what equivalent
families of edge orientations become. Two line arrangements with $D$
indexed lines will be equivalent if they only differ by a
re-indexation of the lines and continuous geometric transformations on
the lines which do not create triple points, because a triple point
corresponds to three coplanar vectors $\evect{a}$, $\evect{b}$,
$\evect{c}$. The equivalence classes of line arrangements in the
projective plane, from $4$ to $7$ lines, are given by
Gr\"unbaum~\cite{grunbaum_b67} and are displayed in
figure~\ref{f_linesarrangement}.  We use in the following the
indexation of line arrangements of this figure
\ref{f_linesarrangement}.  There exists only one arrangement of four
and five lines, whereas there exist four arrangements of six lines and
eleven arrangements of seven lines, which correspond to as many
different $4\ra3$, $5\ra3$, $6\ra3$, and $7\ra3$ random tiling
problems. The icosahedral symmetry belongs to the equivalence class of
the first arrangement of six lines.

\subsection{Polyhedral boundaries}

In the following, we study the presence of cycles in tilings associated with
all those line arrangements. But let us first discuss how those line
arrangements are related to the boundary of the tilings we are
considering. 

\begin{figure}
\includegraphics[width=5.5cm]{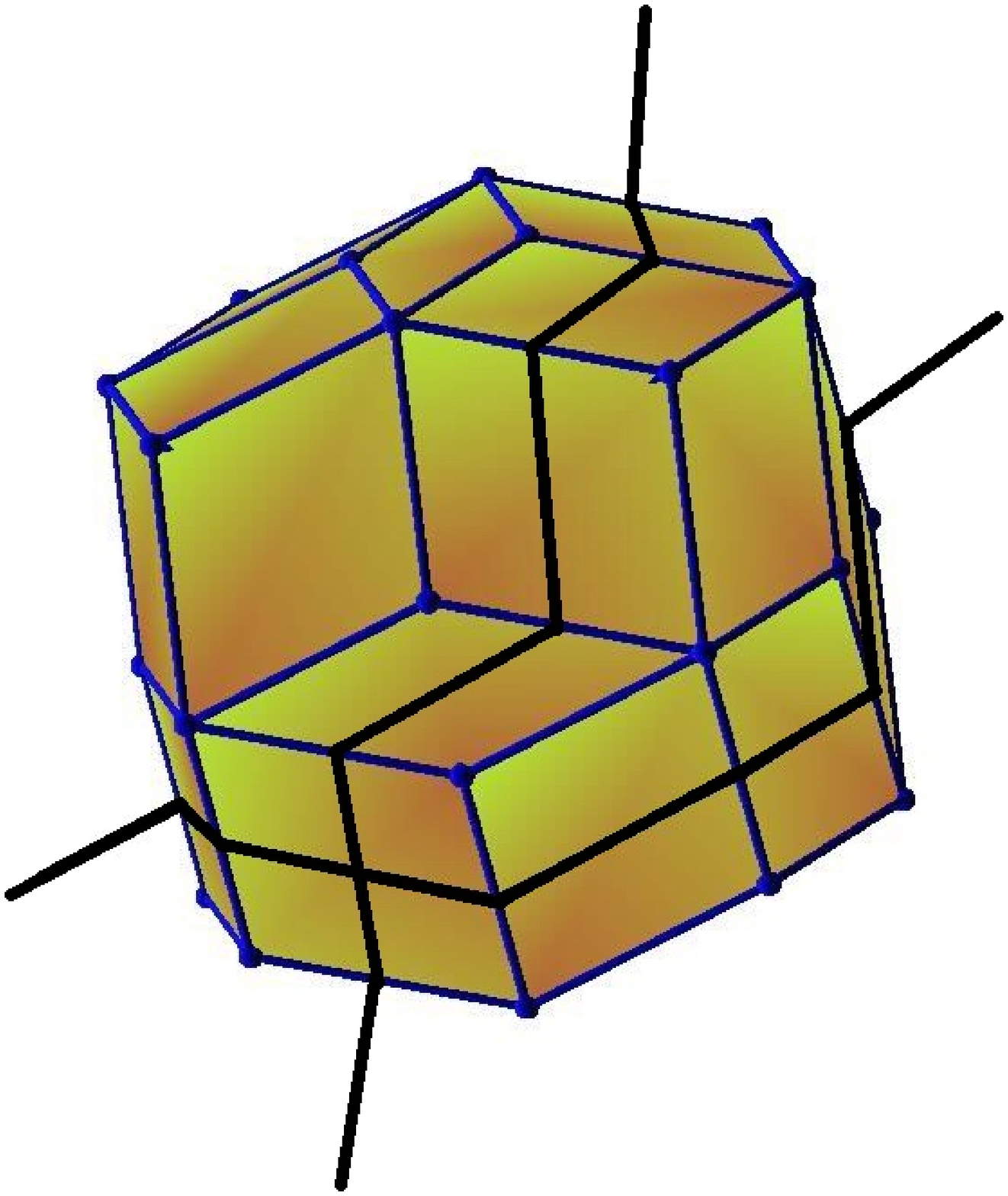}
\hfill
\includegraphics[width=5.5cm]{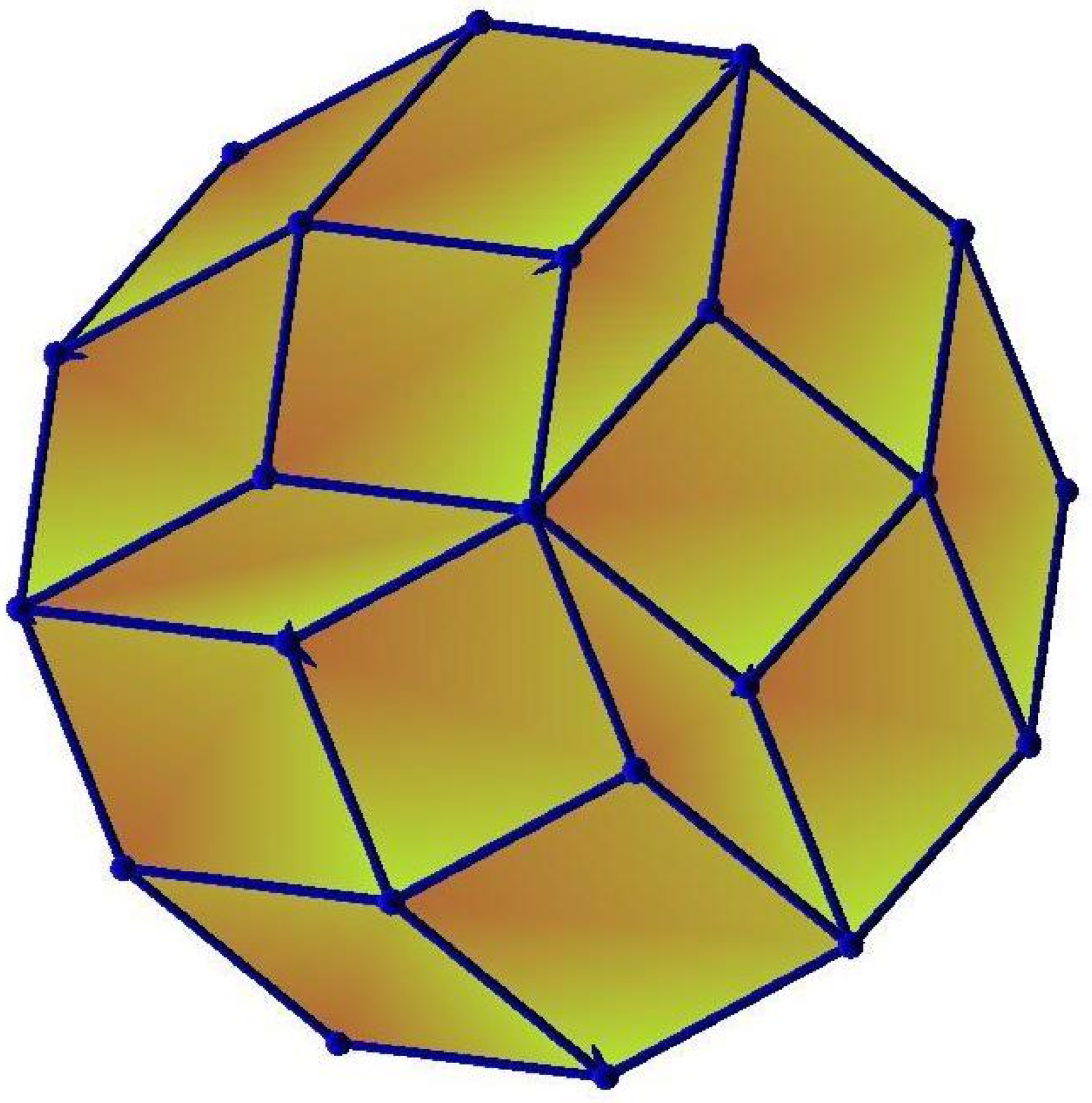}
\caption[]{Two topologically different boundaries of (unitary) $6\ra3$
tilings. The left one corresponds to the icosahedral symmetry
and the right one to the fourth arrangement of six lines.  On
the left one, we have represented the trace of two de Bruijn surfaces
on the boundary by two black lines. By drawing the lines corresponding
to all the de Bruijn surfaces on this boundary, one builds an
arrangement of lines which can be stretched and one obtains the first
6-line arrangement.  A visible difference between these two boundaries
is the connectivity of each vertex.  On the right boundary there
exists a vertex with six edges which is absent on the left one.  This
vertex of connectivity six is represented by an irregular hexagon in
the fourth 6-line arrangement.  }
\label{f_boundarylines}
\end{figure}

The boundary of the tiling generated by the partition-on-tiling method
is the boundary of the Minkowski sum of the vectors $\evect{a}$:
\begin{equation}
Z = \{\sum_{a=1}^{D} \alpha_a \evect{a}, \ \alpha_a \in \Rb, \ 0\leq
\alpha_a\leq p_a \}, 
\end{equation}
which is also called the {\it zonotope} generated by the vectors
$(\evect{1},\evect{2},\ldots, \evect{D})$~\cite{coxeter,destain97} (in
dimension 2, this zonotope is always a $2D$-gon of sides
$(p_1,\dots,p_D)$, which is reminiscent of the uniqueness of edge
orientations). Zonotopes are convex and centro-symmetric. This
boundary is uniquely determined by the choice of the edge
configurations, and thus by the choice of the line arrangement in
dimension 3.  One can directly see this line arrangement by seeing one
hemisphere of the boundary of a unitary tiling projected on a plane
(see figure~\ref{f_boundarylines}). The corresponding line arrangement
is made by the line crossing each family of edges. The projection of
the boundary hemisphere can be seen as a $D \ra 2$ tiling and the
previous line arrangement can be seen as its de Bruijn grid. These
lines are not straight but they can be stretched without changing the
crossing topology.  Indeed this line arrangement can also be seen as
the trace on the projective plane of de Bruijn grid dual of a $D \ra
3$ tiling filling the zonotope. By definition, the boundary does not
depend on the tiling inside this boundary. The line arrangement
corresponding to the boundary can always be seen as the trace of the
dual de Bruijn grid made by flat de Bruijn surfaces, which always
represents a possible tiling. In the figure~\ref{f_boundarylines}, one
can see two tiling boundaries corresponding to the icosahedral tiling
and to the fourth line arrangement with six lines. In particular these
two boundaries differ by the existence of a vertex of connectivity six
in the right-hand-side one. This is seen in the dual fourth line
arrangement by the presence of one hexagon. By contrast, the total
number of tiles $N_T$ does not depend on the choice of the boundary
because the de Bruijn grid is complete:
\begin{equation}
N_T = \sum_{a<b<c} p_a p_b p_c.
\end{equation}

To characterize the differences between random tilings with
non-equivalent edge orientations, we have enumerated tilings with
unitary boundaries ({\it i.e.} with one de Bruijn surface per family),
see tables~\ref{t_entropy63} and \ref{t_entropy73} . For unitary
tilings it is possible to span all the configuration space for each
boundary condition. Indeed, we demonstrate in the following that the
configuration space of all the random tilings with {\em unitary}
boundaries are connected for codimensions up to $4$ which are under
interest in the present paper.  These results, which are exact, show
definitively that tiling problems with non-equivalent families of edge
orientations cannot be put in one-to-one correspondence since they do
not have the same number of configurations. There are 160 unitary
tilings built on the first 6-line arrangement, that is to say with
icosahedral symmetry.

\begin{table}[h]
\resizebox{\textwidth}{!}{
\begin{tabular}{|p{5cm}||*{4}{c|}} 
\hline 
\multicolumn{5}{|c|}{\rule[-3pt]{0pt}{15pt}\large\bfseries 
						 Unitary $6 \rightarrow 3$ tiling space
} \tabularnewline \hline 
Line arrangement & 1 & 2 & 3 & 4 \tabularnewline \hline 
$\sharp$ Tilings	& 160& 148& 144& 148\tabularnewline \hline 
\end{tabular} 

} 
\caption[]{Number of tilings in the
configuration spaces of unitary $6\ra3$ tilings as a function of the
line arrangement. These results have been computed by entirely
spanning the configuration space numerically.}
\label{t_entropy63}
\end{table}

\vspace{0.3cm}

\begin{table}[h]
\resizebox{\textwidth}{!}{
\begin{tabular}{|p{3cm}||*{11}{c|}} 
\hline 
\multicolumn{12}{|c|}{\rule[-3pt]{0pt}{15pt}\large\bfseries 
						 Unitary $7 \rightarrow 3$ tiling space
} \tabularnewline \hline 
Line arrangement 
& 1 & 2 & 3 & 4 & 5 & 6 & 7 & 8 & 9 & 10 & 11 \tabularnewline \hline 
$\sharp$ Tilings	& 7686& 8260& 7624& 7468& 7220& 7690& 7518& 7242& 7106& 6932& 6902\tabularnewline \hline 
\end{tabular} 

}
\caption[]{Number of tilings in the
configuration spaces of unitary $7\ra3$ tilings as a function of the
line arrangement.}
\label{t_entropy73}
\end{table}

\section{Connectivity and structure of the configuration space}
\label{connex}

This section contains results concerning the configuration space of
3-dimensional tilings. First we formulate our main theorem
in~\ref{connex_proof}. It gives a general sufficient condition to
prove that base tilings are acyclic. We apply this result to
dimension 3 in~\ref{dim3} to prove that most line arrangements in
codimension up to 4 generate connected tiling sets. Then we considerer
in~\ref{MFT} the opposite situation where cycles exist and we study
how abundant they are. A mean-field argument supported by numerical
simulations shows that, in this case, cyclic tilings are generic and
acyclic ones are exceptional at the large size limit. Then
in~\ref{struct} we characterize the structure of the configuration
space in the frame of order theory.

\subsection{Connectivity: Main theorem}
\label{connex_proof}

The results of this subsection are not specific to dimension 3 and can
easily be adapted to any $D \ra d$ tiling problem. However, for sake
of simplicity, we shall write our theorem and its proof in dimension
3. Our goal is to give a sufficient condition to prove that base
tilings are acyclic.

We consider a $D+1 \ra 3$ tiling problem defined by a family
$f=(\evect{1},\ldots,\evect{D+1})$ of edge orientations. The set of
all the possible tilings of this tiling problem is denoted by $\TC$.
Tilings of $\TC$ are coded by generalized partitions on base tilings
defined by the $D$ first edge orientations and the faces of which are
oriented by $\evect{D+1}$. The set of base tilings is denoted by
$\tilde{\TC}$. To formulate our main theorem, we first need
introducing some new notions.

First of all, we say that the family
$f=(\evect{1},\ldots,\evect{D+1})$ is {\em acyclic} if any $D \ra 3$
base tiling (of any size) of edges
$\tilde{f}=(\evect{1},\ldots,\evect{D})$, the faces of which are
oriented by $\evect{D+1}$, is acyclic.

A family of $D$ indices $(q_a)$, $a=1,\ldots,D$, is attached to each
tile $u$ of a base $D \ra 3$ tiling $\tilde{t}$ as follows. For each
family of de Bruijn surfaces $F_a$, if $u$ belongs to a surface $S_k$
of $F_a$, then the index $q_a(u)$ is equal to $k$. If $u$ lies between
the surfaces $S_k$ and $S_{k+1}$, in the domain $D_k$, then $q_a(u)$
is half-integer and is equal to $k+1/2$. Let us remark, that when we
code a $D+1 \ra d$ tiling, $t$, by a generalized partition on
$\tilde{t}$, the parts $X_u$ on each tile of $\tilde{t}$ correspond to
the domain, defined by the family $F_{D+1}$, they belong
to: For the tiles $u$ of $\tilde{t}$ in $t$, we have $q_{D+1}(u) = X_u
+ 1/2$.

For each $a$, $q_a$ defines a function on the partially ordered
set of the tiles. We say that $q_a$ is {\em
monotonous increasing} if given any two tiles $u$ and $v$, if $u \leq v$ then
$q_a(u) \leq q_a(v)$. Note that we can in a similar way define the
notion of {\em monotonous decreasing} $q_a$, which is equivalent to
the previous one up to a sign reversal of $\evect{a}$.

We also define {\em companion vectors} in the family $f$: any two
vectors $\evect{a}$ and $\evect{b}$ in $f$ are said to be {\em
companion} if either $\evect{a}$ and $\evect{b}$ orient equivalently
all the faces made by the remaining vectors of $f-\{
\evect{a},\evect{b} \}$, or $\evect{a}$ and $-\evect{b}$ do.

We shall prove below the following lemma:

\medskip

\noindent {\bf Lemma:} {\em if $\evect{a}$ and $\evect{D+1}$ are
companion vectors in the family of edge orientations $f$, then the
function $u \mapsto q_a(u)$ is monotonous (increasing or decreasing).}

\medskip

As a consequence, in this case, $q_a$ cannot but be constant along a
cycle.  Therefore a cycle either lies entirely in a single de Bruijn
surface of $F_a$ (if $q_a$ is an integer) or strictly lies between two
of them (if $q_a$ is half-integer). In the first case, the cycle lives
in the equivalent of a $D-1 \ra 2$ tiling (see section~\ref{gene}),
and in the second case in a $D-1 \ra 3$ tiling of edge orientations
$(\evect{1},\ldots,\evect{a-1},\evect{a+1},\ldots,\evect{D})$. We
already know that cycles do not exist in any $D-1 \ra 2$
tiling. Therefore if we can prove inductively that cycles cannot exist
in the $D-1 \ra 3$ tilings (in other words that $f-\{\evect{a}\}$ is
acyclic), we get that they cannot exist in base $D \ra d$ tilings under
interest. We are led to our main theorem:

\medskip

\noindent {\bf Main theorem:} {\em Given a family of $D+1$ 3-dimensional
edge orientations $f=(\evect{1},\ldots,\evect{D+1})$, if there
exists a vector $\evect{a}$ in $f$ companion of $\evect{D+1}$,
and if the family $f-\{\evect{a}\}$ is acyclic, then $f$ is acyclic
in its turn. }

\medskip

This theorem will be used in the following subsection as follows: we
shall prove that some families of edges are acyclic, proceeding by
induction on their numbers of vectors. Connectivity of all fibers in
the fibration corresponding to $\evect{D+1}$ will follow. If in
addition we know that the set $\tilde{\TC}$ of base tilings is
connected, we shall get the connectivity of the whole set of tilings
$\TC$.

To finish with, we prove our first lemma. We prove that if $\evect{a}$
and $\evect{D+1}$ orient equivalently all the faces made by the
remaining vectors, then $q_a$ is monotonous {\em increasing}. We
could prove in a similar way that if $\evect{a}$ and $-\evect{D+1}$
orient equivalently all the faces made by the remaining vectors, then
$q_a$ is monotonous {\em decreasing}.

To establish the monotony of $q_a$, on all the tilings based on the
family $f-\{\evect{D+1}\}$, we consider two adjacent tiles $u$ and $v$
such that $u \leq v$ ({\em i.e.} $u \leq_{D+1} v$). Three cases may
occur: (i) $u$ and $v$ belong to two different de Bruijn surfaces of
family $F_a$.  Since $u \leq v$, and $\evect{a}$ companion of
$\evect{D+1}$, $v$ is above $u$ along the direction $\evect{a}$, then
$q_a(v)=q_a(u)+1$; (ii) $u$ belongs to a de Bruijn surface of $F_a$
and $v$ belongs to a domain $D_k$ between two such surfaces; or $u$
belongs to such a domain and $v$ belongs to a de Bruijn surface.  In
both cases, for the same reasons as in (i), $q_a(v)=q_a(u)+1/2$; (iii)
$u$ and $v$ belong to the same domain $D_k$ or to the same de Bruijn
surface $S_k$: $q_a(v)=q_a(u)$. Therefore, in all cases, $q_a(u) \leq
q_a(v)$, which proves monotony.

\subsection{Connectivity in dimension 3}
\label{dim3}

We now exhibit a criterion to identify companion vectors in
the line arrangements in the projective plane corresponding 
to families of 3-dimensional edge orientations. Since 
any vector in $f$ can {\em a priori} play the role of $\evect{D+1}$, we
seek any two companion vectors in $f$.

In order to find companion vectors, we represent the projective plane by
the unit 2-sphere $S^2$ on which antipodal points are identified. The
lines on $\Pb\Rb^2$ are now represented by great circles $\Gamma_a$ on
$S^2$, such that $\Gamma_a$ lies in a plane perpendicular to
$\evect{a}$. This circle separates $S^2$ into two hemispheres. We
define a positive hemisphere $\Gamma_a^+$ and a negative
one $\Gamma_a^-$ such that $\evect{a}$ points from $\Gamma_a^-$ to
$\Gamma_a^+ $.

Here we denote by $\FC_{cd}$ the face species that is defined by the
vectors $\evect{c}$ and $\evect{d}$. It is represented in $S^2$ by the
two intersections of the circles $\Gamma_c$ and $\Gamma_d$, denoted by
$\gamma_{cd}^{(1)}$ and $\gamma_{cd}^{(2)}$. These points are
identified in $\Pb\Rb^2$.

Now we prove that a face $\FC_{cd}$, where $c$ and $d$ are different
from $a$ and $b$, is oriented equivalently by $\evect{a}$ and
$\evect{b}$ if and only if $\gamma_{cd}^{(1,2)}$ belongs to $\Gamma_a^+
\cap \Gamma_b^+$ or to $\Gamma_a^- \cap \Gamma_b^-$.

We assign the orientation of $\FC_{cd}$ by $\evect{a}$, by a unitary
vector $\vect{n}_{cd}$, normal to $\FC_{cd}$, and crossing $\FC_{cd}$
in the same direction as $\evect{a}$, that is to say $\vect{n}_{cd}
\cdot \evect{a} > 0$. In other words, $\vect{n}_{cd} \in \Gamma_a^+$.
We obtain in the same way that $\vect{n}_{cd} \in \Gamma_b^+$, since
$\evect{a}$ and $\evect{b}$ orient equivalently $\FC_{cd}$.  Now, by
definition, $\vect{n}_{cd}$ is unitary and is perpendicular both
to $\evect{c}$ and $\evect{d}$. Thus it coincides with $\gamma_{cd}^{(1)}$ or
$\gamma_{cd}^{(2)}$. Hence $\gamma_{cd}^{(1,2)}$ belongs to
$\Gamma_a^+ \cap \Gamma_b^+$ or to $\Gamma_a^- \cap \Gamma_b^-$.

Therefore two vectors $\evect{a}$ and $\evect{b}$ are companion if and
only if all the points $\gamma_{cd}^{(1,2)}$ with $c$ and $d$
different from $a$ and $b$ belongs to $(\Gamma_a^+ \cap \Gamma_b^+)
\cup (\Gamma_a^- \cap \Gamma_b^-)$, or all of them belong to
$(\Gamma_a^+ \cap \Gamma_b^-) \cup (\Gamma_a^- \cap \Gamma_b^+)$
(in the case where it is $\evect{a}$ and $-\evect{b}$ which
orient equivalently all the faces).

\begin{figure}[h!]
\includegraphics[width=6cm]{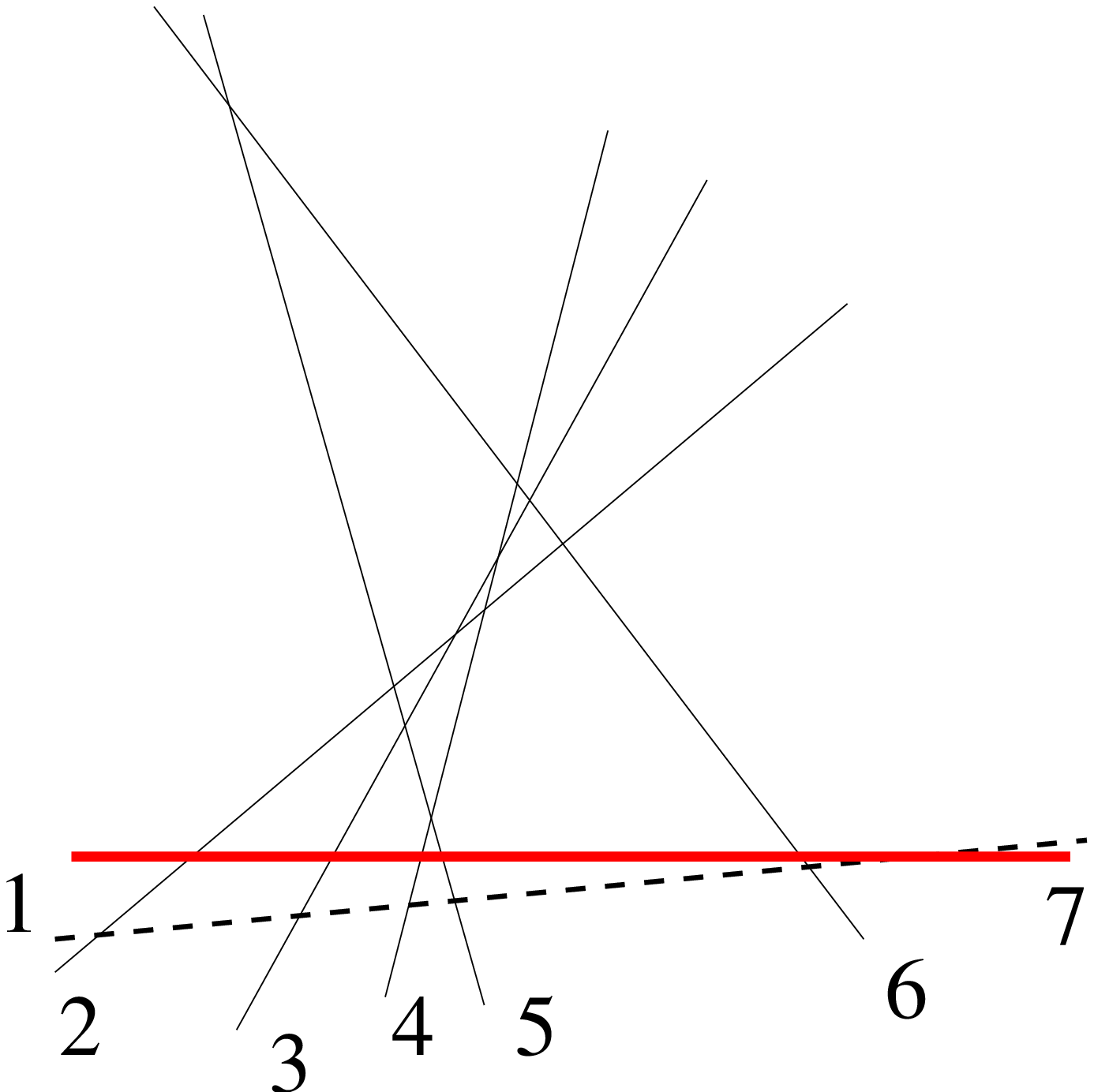}~\hfill~\includegraphics[width=8cm,height=6.5cm]{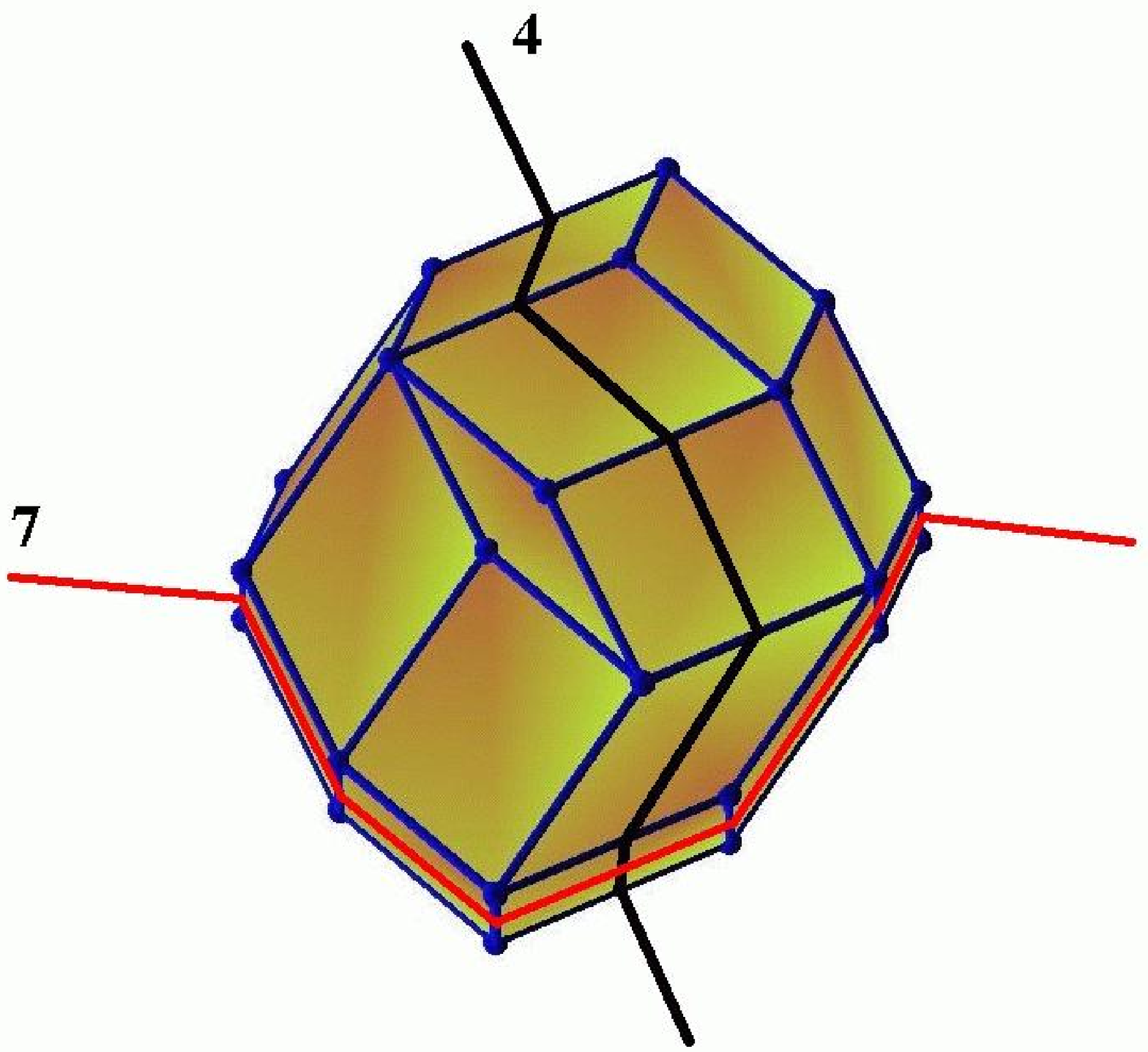}
\caption[]{On the left, the tenth 7-line arrangement where we have
represented two companion lines, $1$ and $7$. All the intersections of
any two remaining lines are on the same sides of these companion
lines. On the right, one possible view of the corresponding boundary
of this line arrangement. The vector $\evect{1}$ (which orients the
base tilings in this figure) is perpendicular to the plane of the
figure and points upward. The corresponding tiling edges cannot be
seen on this figure. But one can easily see that $\evect{7}$ is
companion to $\evect{1}$: The trace of the seventh de Bruijn surface
is represented on this boundary by the bottom curve.  It crosses
edges borne by $\evect{7}$. One can check that, because of convexity
of the zonotopal boundary, the sign of $\evect{7}$ can be chosen so as
to orient all the faces which do not belong to the trace of the
seventh de Bruijn surface from bottom to top as $\evect{1}$ does.  On the
contrary, the vector $\evect{4}$ is not companion of $\evect{1}$: The
trace of the forth de Bruijn surface is represented by a line which
splits the boundary into tow non-empty parts. If the faces on the left
of this line are oriented from bottom to top by $\evect{4}$, it will
orient the ones on the right from top to bottom.}
\label{f_boundary_companion}
\end{figure}

In practice to find companion vectors one can use any hemisphere of
$S^2$ represented by an affine plane to which we add the line at
infinity. If this plane can be chosen so that all the points made by
the intersection of two lines different from $\LC_a$ and $\LC_b$ are
on the same sides of $\LC_c$ and $\LC_d$, then $\evect{c}$ and
$\evect{d}$ are companion vectors. An example is displayed in
figure~\ref{f_boundary_companion} (left).

The previous analysis can also be understood geometrically by
considering one arbitrary hemisphere of the boundary of a unitary
tiling corresponding to a given arrangement, as displayed in
figure~\ref{f_boundary_companion} (right). In this figure, the vector
$\evect{D+1}$ is perpendicular to the figure plane. It orients all the
faces from bottom to top. Since the zonotope is convex, one can easily
check that in this representation, a companion $\evect{a}$ of this
vector is such that the arrangement is completely situated on one side
of the line $\LC_a$. Whereas a vector $\evect{b}$ such that the line
$\LC_b$ divides the arrangement into two non-empty parts cannot be a
companion of $\evect{D+1}$.  Indeed, if $\evect{b}$ orients the faces
on the right of $\LC_b$ from bottom to top, one can see that it
orients the faces on the left of $\LC_b$ from top to bottom.

We have systematically applied this criterion on the line arrangements
of figure~\ref{f_linesarrangement} to identify companion vectors, in
order to apply inductively our main theorem. Tables \ref{t_fitcycle53}
and \ref{t_fitcycle63} provide a summary of our investigations.  For
each line arrangement, in which the particularized line
$\lambda_{D+1}$ corresponds to the vector
$\evect{D+1}$ orienting base tilings, a 0 indicates that the
corresponding family of edge orientations is acyclic.  If for a given
arrangement, there exists a $\lambda_{D+1}$ for
which it is the case, we can conclude that the configuration space is
connected after checking that the base is itself connected.

Conversely, there are arrangements for which we cannot find any pair
of companion lines. The first arrangement of six lines is an example.
It means that we cannot state on the absence of cycles in any
generalized partition based on this arrangement. Actually, we will see
in the following that we {\em do} find cycles on these generalized
partitions. In fact, for each arrangements for which we cannot prove
the absence of cycles we find them by numerical exploration, see
tables~\ref{t_fitcycle53} and \ref{t_fitcycle63}, and section
\ref{MFT}. There exist only three arrangements (among 17) of at most
seven lines for which there is no fibration without cycles and the
connectivity of which remains open, as it is displayed in the
tables. So for all the remaining 14 arrangements, we prove the
connectivity of the corresponding tiling sets for any tiling size.
 
By contrast, sets of {\em unitary} tilings are connected for any
codimension lesser than 4. Indeed, we proved by systematic numerical
exploration of sets of unitary tilings of codimension lesser than 3 that
there always exists an acyclic fibration in this case.

In order to prove the connectivity for the tiling problems for which
all fibrations exhibit cycles, we would have to prove that one can
always break the cycles by type-I flips (see
figure~\ref{f_cycle_example}). Which means that there exists a
sequence of flips which brings the tiling from any disconnected fibers
to a connected one. Then one can change the parts of the tiles of
previous cycles by type II-flips, and bring back the tiling to another
component of the initial disconnected fiber. We have not established a
possibility to break those cycles in all cases, and therefore the
connectivity problem remains open.

In principle, the arguments developed in this section can be extended
to any $D \ra d$ tiling problem using arrangements of hyper-planes in
the projective space $\Pb\Rb^{d-1}$ provided one knows their
classification for fixed $D$ and $d$. In \ref{codim2}, we prove
connectivity for codimension 2 tilings of any dimension. Note that
independently of this work, and after introducing a new and different
formalism, Fr\'ed\'eric Chavanon and \'Eric R\'emila quite recently
also established connectivity in codimension 2~\cite{chavanon04}.

\subsection{Abundance of cyclic base tilings~; mean-field argument 
and numerical studies}
\label{MFT}

We now present numerical results on the abundance of cyclic base
tilings of dimension 3 as a function of their boundary size, as well
as a mean-field argument to account for these results. In this
section, we focus on diagonal tilings: $p_a=p$ for any $a$. For
unitary tilings, we completely cover the configuration space (by use
of a deep search algorithm), therefore the results are exact. For
larger tilings, we numerically sample the tiling configuration
space by use of the Monte Carlo Markovian dynamics described in
section~\ref{s_cycle}. 

\begin{figure}[h!]
\begin{center} 
\rotatebox{-90}{\resizebox{7cm}{!}{\includegraphics{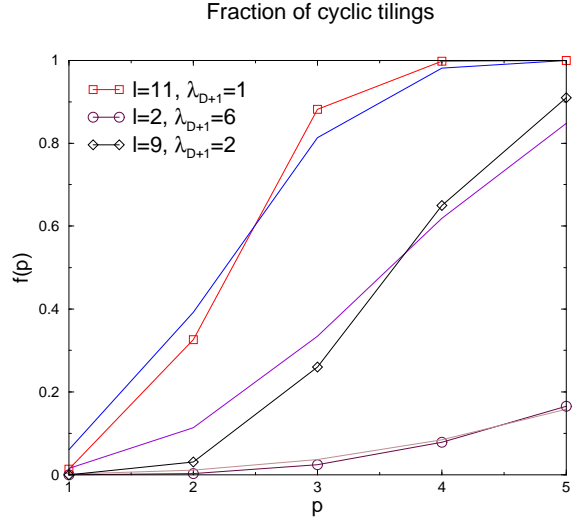}}} \\
\end{center}
\caption[]{Typical fractions of cyclic tilings (when they are not equal
to zero) as a function of boundary size. The curves with squares,
diamond and circles are the measured ones, whereas the others are fits
with: $1-(1-\alpha)^{20p^3}$. We represent only some cycle
fractions for $6\ra3$ tilings for visibility.}
\label{f_fracycle}
\end{figure}

We have made this numerical investigation for all the tiling problems
corresponding to 6 and 7-line arrangements, with one particularized
line of index $\lambda_{D+1}$ corresponding to $\evect{D+1}$ to orient the
base tilings. Typical cycle abundance we found are displayed
in figure~\ref{f_fracycle}.  For all the tiling problems for which we
prove that cycles cannot exist we effectively never find them.  For
the other ones, we find cycles and their occurrence increases rapidly
with the tiling size $p$ for all cases with a similar law. The
differences come from that cycle existence does not arise at the same
size for all those tiling problems. In particular, cycles do not exist
in unitary $5\ra3$ tilings (exact result) and they appear starting
from $p=2$. For $6\ra3$ tilings for which there exist cycles in the
unitary case, the cycle fraction is already close to $1$ for
$p=5$. These results show that when cycles can exist in a tiling
problem, they are certainly very frequent for tilings at large size
$p$. Indeed, if a small cycle appears in a small tiling, it will be
likely to appear locally in a large one which can be seen in a first
approximation as a juxtaposition of nearly independent smaller
tilings.

Following this idea, we now propose a mean-field argument to account 
for these results. For a tiling problem in which cycles are possible, we
suppose that there is a non-zero probability $\alpha$ that a tile
belongs to a cycle. If all the tiles are considered as independent,
the probability that no tile belongs to a cycle in the whole tiling is
then $(1-\alpha)^{N_T}$ and the fraction of cyclic tilings reads:
\begin{equation}
 1-(1 -\alpha)^{N_T}. 
\end{equation}
This supposes that $\alpha$ is independent of the tiling size. It is
necessarily false since we have seen that in some cases cycles appear
only starting from a given size. But one can think that it is a good
approximation for large sizes. The fits of the measured fraction of
cyclic tilings by this simple law reproduce correctly the results in all
the cases, see figure~\ref{f_fracycle}. A summary of these fits is
given in tables~\ref{t_fitcycle53} and \ref{t_fitcycle63}. We mention
that tiling problems based on the same line arrangement but with a
different particularized line $\lambda_{D+1}$ are not necessarily
different. In particular, the 6 fibrations of the $6\ra3$ tiling
problem corresponding to the first 6-line arrangement are all
equivalent because all 6 lines play the same role with respect to
the 5 remaining ones. They are represented by the same $\alpha$ in
table~\ref{t_fitcycle53}.
    
Concerning the connectivity problem one can see that it remains only
three open cases. One of them corresponds to the first arrangement of
6 lines and so to the icosahedral symmetry. Note that it can bias the
issue of few results on the fractions of cyclic base $6\ra3$ tilings
in table~\ref{t_fitcycle63}. Indeed, the corresponding base tiling
sets might be disconnected and our Monte Carlo sampling might be
incorrect.  These 3 concerned cases are indicated in bold faces in
the table.

\begin{table}[h!]
\resizebox{\textwidth}{!}{
\renewcommand{\multirowsetup}{\centering} 
\renewcommand{\bigstrutjot}{1pt} 
\begin{tabular}{|c||c||*{6}{c|}c|} 
\hline 
\multicolumn{9}{|c|}{\rule[-5pt]{0pt}{20pt}\large\bfseries 
					Cycle abundance in
$5\rightarrow 3$ base tilings fitted by $1-(1-\alpha)^{{5 \choose 3}p^3}$} \tabularnewline  \hline 
  & 

\multirow{2}{2.2cm}{\rule[-0pt]{0pt}{7pt} 6-line \rule[-0pt]{0pt}{7pt} arrangement \rule[-0pt]{0pt}{7pt}number}

  & \multicolumn{6}{|c|}{ \rule[-5pt]{0pt}{20pt}Particularized line 
$\lambda_{D+1}$} & 

\multirow{2}{2.2cm}{\rule[-0pt]{0pt}{7pt} Connectivity problem }

\tabularnewline \cline{3-8} \bigstrut[b]

  \rule[-2pt]{0pt}{20pt}&  \rule[-2pt]{0pt}{20pt}& \rule[-2pt]{0pt}{20pt} {\bf 1} 

&\rule[-2pt]{0pt}{20pt}  {\bf 2} & \rule[-2pt]{0pt}{20pt} {\bf 3} 

& \rule[-2pt]{0pt}{20pt} {\bf 4} &\rule[-2pt]{0pt}{20pt}  {\bf 5} 

&\rule[-2pt]{0pt}{20pt}  {\bf 6} &\rule[-2pt]{0pt}{20pt}

\tabularnewline  \hline \hline  
$\alpha$  & {\bf1}
& 1.32e-06 & 1.32e-06 & 1.32e-06 & 1.32e-06 & 1.32e-06 & 1.32e-06 & open\tabularnewline \hline 
$\alpha$  & {\bf2}
& 0& 0& 0& 0& 0& 0 & connected\tabularnewline  \hline 
$\alpha$  & {\bf3}
& 0& 0& 0& 0& 0& 0 & connected\tabularnewline  \hline 
$\alpha$  & {\bf4}
& 0& 0& 0& 0& 0& 0 & connected\tabularnewline  \hline 
\end{tabular} 

} 
\caption[]{Results of the fits of fractions of cyclic tilings by
$1-(1-\alpha)^{10p^3}$. The 0 represents fibrations where we 
proved that cycles cannot appear. The connectivity problems remains
open for the first arrangement (icosahedral symmetry). }
\label{t_fitcycle53}
\end{table}


\begin{table}[h!]
\resizebox{\textwidth}{!}{
\renewcommand{\multirowsetup}{\centering} 
\renewcommand{\bigstrutjot}{1pt} 
\begin{tabular}{|c||c||*{7}{c|}c|} 
\hline 
\multicolumn{10}{|c|}{\rule[-5pt]{0pt}{20pt}\large\bfseries 
					Cycle abundance in
$6 \rightarrow 3$ base tilings fitted by $1-(1-\alpha)^{{6 \choose 3}p^3}$} \tabularnewline  \hline 
  & 

\multirow{2}{3cm}{\rule[-0pt]{0pt}{7pt} 7-line \rule[-0pt]{0pt}{7pt} arrangement  \rule[-0pt]{0pt}{7pt} number}

  & \multicolumn{7}{|c|}{ \rule[-5pt]{0pt}{20pt} Particularized line $\lambda_{D+1}$}  & 

\multirow{2}{2.2cm}{\rule[-0pt]{0pt}{7pt} Connectivity problem }

\tabularnewline \cline{3-9} \bigstrut[b]

  \rule[-2pt]{0pt}{20pt}&  \rule[-2pt]{0pt}{20pt}& \rule[-2pt]{0pt}{20pt} {\bf 1} 

&\rule[-2pt]{0pt}{20pt}  {\bf 2} & \rule[-2pt]{0pt}{20pt} {\bf 3} 

& \rule[-2pt]{0pt}{20pt} {\bf 4} &\rule[-2pt]{0pt}{20pt}  {\bf 5} 

&\rule[-2pt]{0pt}{20pt}  {\bf 6} &\rule[-2pt]{0pt}{20pt} {\bf 7} & \rule[-2pt]{0pt}{20pt}

\tabularnewline  \hline \hline  
$\alpha$ & {\bf1} 
& 0& 0& 0& 0& 0& 0& 0 & connected\tabularnewline \hline 
$\alpha$ & {\bf2} 
& {\bf N.A} & 2.98e-5& 3.57e-5& 2.15e-5& {\bf N.A} & 1.38e-4& 1.13e-4 & open\tabularnewline \hline 
$\alpha$ & {\bf3} 
& 0& 0& 0& 7.58e-3& 0& 0& 0 & connected\tabularnewline  \hline 
$\alpha$ & {\bf4} 
& 0& 0& 0& 0& 0& 0& 0 & connected\tabularnewline \hline 
$\alpha$ & {\bf5} 
& 0& 0& 0& 0& 0& 0& 0 & connected\tabularnewline \hline 
$\alpha$ & {\bf6} 
& {\bf 5.93e-3}& 1.39e-4& 3.32e-4& 2.81e-4& 1.65e-4& 1.93e-4& 3.18e-4 & open\tabularnewline  \hline 
$\alpha$ & {\bf7}
& 0& 0& 0& 2.13e-4& 4.05e-4& 2.14e-4& 0 & connected\tabularnewline \hline 
$\alpha$ & {\bf8} 
& 5.01e-4& 0& 0& 0& 0& 0& 0 & connected\tabularnewline \hline 
$\alpha$ & {\bf9} 
& 0& 1.51e-3& 0& 0& 6.86e-3 & 6.14e-3& 0 & connected\tabularnewline  \hline 
$\alpha$ & {\bf10} 
& 0& 0& 0& 0& 6.07e-3 & 0& 0 & connected\tabularnewline  \hline 
$\alpha$ & {\bf11} 
& 6.20e-3& 0& 0& 0& 0& 0& 0 & connected\tabularnewline \hline 
\end{tabular} 

}
\caption[]{Results of the fits of fractions of cyclic tilings by
$1-(1-\alpha)^{20 p^3}$. The acronym ``N.A'' holds for non-available: in
these cases, we found cycles but not enough for the reliability of the
fit.  The connectivity problem remains open only for the second and the
sixth line arrangements, since for the others we can prove that at least
one fibration cannot possess cycles. All configuration sets of
$6 \ra 3$ {\em base tiling problems} are proven to be connected 
except the 3 ones written in bold faces.}
\label{t_fitcycle63}
\end{table} 

\subsection{Structure of the configuration space}
\label{struct}

We now make a brief incursion into graph theory and order theory.  The
configuration space can be seen as a graph $G$, the vertices of which
represent tilings, and the edges of which represent single flips:
given two tilings $t_1$ and $t_2$, $(t_1,t_2)$ is an edge of $G$ if
$t_1$ and $t_2$ differ by (only) one single flip (see
figure~\ref{f_space_fiber}). We prove that this graph $G$ can be
embedded in a high-dimensional hyper-cubic lattice $L$, thus
generalizing results previously specialized to plane octagonal
tilings~\cite{octo01}, even in the possible case where $G$ is not
connected.

As an immediate corollary, we demonstrate that this graph can be given
a structure of {\em graded partially ordered set} (graded
``poset'')~\cite{davey02}. Indeed a partial order relation is
associated below to the iterated partition-on-tiling process. Saying
that this poset is graded means that there exists a {\em rank
function} $r$ on configurations such that if $t_1$ covers ({\em i.e.} is
just above) $t_2$ then $r(t_1) = r(t_2)+1$. This rank function is simply the
sum of the (integral) coordinates of a tiling in the lattice $L$. This
order has unique minimal and maximal elements.

The present point of view is applicable to tiling sets of any
dimension $d$ and codimension $D-d$.

\subsubsection{Structure of the graph \\}

We first prove that the configuration space can be seen as a graph
$G_D$ embedded in a high-dimensional hyper-cubic lattice $L_D$. We proceed by
induction on the codimension $D-d$, for any fixed dimension $d$.

In codimension 1, the tilings are encoded by hyper-cubic
partitions. The coordinates of a tiling are simply the $K_{d+1}$ parts
$x_k$, $k=1,\ldots,K_{d+1}$, of its associated partition and a tiling
is therefore naturally represented by a point of integral coordinates
in a hyper-cubic lattice $L_{d+1}$ of dimension $K_{d+1}$. A flip is
encoded by an increase or decrease of the corresponding part by one
unit and thus it corresponds to an edge of the hypercubic lattice.

Suppose now that the above property holds for the graph $G_{D}$.  A
$D+1 \rightarrow d$ tiling $t$ is encoded by both a $D \rightarrow d$
base tiling $\tilde{t}$ and a generalized partition on this base
tiling. By hypothesis, the base tiling is encoded by $K_D$ integral
coordinates $x_k$, $k=1,\ldots,K_D$. We denote by $y_l$,
$l=1,\ldots,K'$ the parts of the partition. We now demonstrate that if
$t$ is encoded by the coordinates
$(x_1,\ldots,x_{K_D},y_1,\ldots,y_{K'})$, then $G_{D+1}$ is embedded
in a lattice $L_{D+1}$ of dimension $K_{D+1}=K_D+K'$.

The only subtlety comes from the fact that the indices $l$ must be
correctly chosen with respect to the base tiling $\tilde{t}$, so
that $G_{D+1}$ is {\em globally} embedded in a hyper-cubic
lattice (and not only {\em locally}), as it is
already discussed in reference~\cite{octo01} in the special case of
octagonal tilings. We do not reproduce the ideas of this reference
which cannot be easily generalized.

A tile of any base tiling $\tilde{t}$ is defined as the intersection
of $d$ de Bruijn surfaces. The de Bruijn families are indexed by
$a_1,\ldots,a_d$ and in each family $a$, the surface is indexed by
$q_a$. A tile is now indexed by the $2d$ indices
$(a_1,\ldots,a_d,q_{a_1},\ldots,q_{a_d})$ independently of
$\tilde{t}$. We simply fix an arbitrary one-to-one correspondence
between these indices and the indices $l=1,\ldots,K'$, independent of
the tiling $\tilde{t}$.

Now that we have defined the integral coordinates of a tiling, we only
need to check that both type-I and type-II flips respect the lattice
structure, that is to say that they correspond to an increase or a
decrease of (only) one coordinate by one unit.

Type-II flips do not affect the base tiling (the $x_k$ are unchanged)
whereas they change exactly one $y_l$ by $\pm 1$; Type-I flips concern
the base tiling only: they change one $x_k$ by $\pm 1$ and $d+1$ tiles
of the base tiling move. Since the flip is possible, they all bear the
same part value before the flip. After the flip, these part values
remain unchanged. Since the tiles involved in the flip bear the same
indices $l$ before and after the flip, the coordinates $y_l$ remain
unchanged. To sum up, either one $x_k$ or one $y_l$ (and only one) is
increased or decreased by one unit when a flip is achieved.

\subsubsection{Structure of graded poset \\}

When $G_D$ is connected by flips, as far as the order structure is
concerned, the previous results ensure that $G_D$ has a structure of
graded poset inherited from the order structure of $L_D$: a tiling
$t_1$ is greater than a tiling $t_2$ if all the coordinates of $t_1$
are greater than the corresponding coordinates of $t_2$ in $L_D$. The
rank function $r(t)$ is simply the sum of the coordinates of $t$ in
the lattice $L_D$. The minimum tiling is obtained when all the parts
are set to 0. The maximum tiling is obtained when all the parts are
set to their maximum possible value. The same kind of result 
is also established in~\cite{chavanon04} in codimension 2.

To close this section, note that the existence of the rank function
$r$ make in principle possible the application of the technique
developed in reference~\cite{arctic} to calculate numerically the entropy
of tilings with a given edge orientation, as soon as the configuration
space is connected by flips.

\section{What about more physical free-boundary tilings?}
\label{free_fixed}

In reference~\cite{arctic}, it is discussed that the fixed boundaries
we consider in the present paper are not physical. The aim of the
present section is to clarify how our results can be transposed to
free- (or periodic-) boundary tilings which are more realistic models
of quasicrystals.  We argue that flip dynamics in fixed-boundary
tilings is relevant to flip dynamics in free-boundary ones provided one
focuses on their central regions, where they forget the
influence of their polyhedral boundary.

Since~\cite{elser84}, it is known that fixed zonotopal boundaries have
a strong influence on rhombus tilings. This boundary sensitivity has
been widely studied in two dimensions (see references
in~\cite{arctic}) and recently numerically explored in $4 \rightarrow
3$ tilings~\cite{Linde01,arctic}. This spectacular effect is
generically known as the ``arctic phenomenon'', which means that at
the large size limit, constraints imposed by the boundary ``freeze''
macroscopic regions near the boundary. In these frozen regions, the
tiling is periodic, contains only one tile species, and has a
vanishing entropy.  By contrast, the remaining ``unfrozen'' regions
contain random tilings with several tile species and have a finite
entropy per tile. In the two-dimensional hexagonal case, the unfrozen
region is inscribed in an ``arctic circle''. The tiling is not
homogeneous inside this circle and presents an entropy gradient.

By contrast, in $4 \rightarrow 3$ tilings filling a rhombic
dodecahedron, its has been numerically established that the unfrozen
region is an octahedron~\cite{Linde01,arctic}, inside which the tiling
is homogeneous and the entropy per tile is constant. In other words,
inside the octahedron, the tiling is a {\em free-}boundary one.
This octahedron contains 2/3 of the tiles.

In~\cite{arctic}, it is argued that this qualitative difference
between two- and three-dimensional tilings is related to the entropic
repulsion between de Bruijn lines and surfaces. In dimension 2, it is
favorable to bend de Bruijn lines because the bending cost is smaller
than the entropy gained by moving lines away.  The reverse holds in
dimension 3 for de Bruijn surfaces: they are not forced away from
their flat configuration and they remain stacked in the octahedron. It
is anticipated in~\cite{arctic} that the same kind of result holds in
higher codimension three-dimensional $D \ra 3$ tilings: if the de
Bruijn surfaces are not forced away from their flat configuration,
there should be a large macroscopic central region where all de Bruijn
families are present, are flat at large scale, and form a
free-boundary $D \ra 3$ tiling, thus forgetting the presence of the
polyhedral boundary. For example, a simple calculation shows that in
diagonal icosahedral tilings, this region contains about 42\% of the
tiles.

In this icosahedral case, locally jammed clusters of tiles associated
with cycles and likely to affect the flip dynamics necessarily belong
to this free-boundary-like central region, because peripheral zones
are of lower codimension and cannot contain cycles.  As a consequence,
these jammed configurations also exist in free- or
periodic-boundary tilings, with all their implications, and are not
specific to fixed boundaries. If they affect the dynamics, it will
certainly also be true in free-boundary tilings, especially in large size ones.

For the same reasons, in the following, we study vertex self-diffusion
in this central region: the initial positions of the vertices are
chosen in a very small central sphere and we check that their distance
to the center never exceeds a finite fraction ($\sim 50$\%) of the
tiling shortest radius. We have argued that we effectively study
self-diffusion in tilings rid of the non-physical influence of fixed
polyhedral boundaries. Below, we compare the diffusion constant in
this central region of fixed-boundary icosahedral tilings with the
similar constant in tilings with periodic
boundaries~\cite{Jaric94}. We find an excellent agreement, which
corroborates that the tiling in the central region is effectively a
free- (or equivalently periodic-) boundary one and which reinforces
our analysis.

\section{Vertex self-diffusion}
\label{diffusion}

In this section, we study vertex self-diffusion in rhombohedra
tilings.  Even though self-diffusion is only one way of characterizing
flip dynamics among many possible ones, we have chosen to focus on
this observable because of its physical interest (see section~\ref{cl}
and the end of this section for a discussion on other quantities of
interest related to flip dynamics).

\subsection{Physical motivation}
\label{physical_diff}

Indeed, single flips have a counterpart at the atomic
level~\cite{coddens93,lyonnard96} which is a new source of atomic
mobility as compared to usual mechanisms in crystals.  Consecutively,
flip-assisted atomic self-diffusion has been anticipated as a
transport process specific to quasi-crystalline
materials~\cite{kalugin93} which is susceptible to play a role in
their mechanical properties, even if it remains controversial whether
or not flip-assisted self-diffusion is dominant as compared to usual
mechanisms~\cite{bluher98}.  

In addition, quasicrystals present a sharp brittle-ductile transition
well below their melting transition (for a review, see Urban {\em et
al.}~\cite{urban99}), which is related to a rapid increase of
dislocation mobility~\cite{wollgarten93}. Note that the latter does
not seem to be directly associated with any phason unlocking
transition because the phason faults dragged behind a moving
dislocation are not healed immediately (neither above or
below the transition) and the friction force on a
dislocation due to the trailing of phason faults should not vary
significantly at the brittle-ductile transition~\cite{caillard_pc}.
However, dislocation movement by pure
climb~\cite{caillard03,mompiou03,caillard_pc} requires the diffusion
of atomic species over large distances. Therefore dislocation mobility
is also directly related to atomic self-diffusion.

Here we study the diffusion of vertices in tilings. As it was first
established in~\cite{kalugin93}, one must focus on diffusion of
vertices rather than diffusion of tiles because tiles cannot travel
long distances under flip sequences. Diffusion of vertices is a first
approximation before a more realistic and refined approach taking into
account atomic decorations of tiles. But the possible effects of
cycles we want to address here are already present at the scale of
tiles and we shall not consider atomic decorations in this paper.  

We demonstrate that cycles do not have any significant influence on
self-diffusion, both at the qualitative and quantitative levels.

\subsection{Numerical results}

Our purpose here is not an exhaustive study of vertex self-diffusion
that was already done in reference~\cite{Jaric94}, but rather to check
that cycles have no significant influence on flip dynamics, at least
as far as vertex diffusion is concerned. We also focus on diagonal
tilings.

We implement our numerical study as follows. We consider the Monte
Carlo Markovian dynamics described in section \ref{s_cycle}.  The unit
of time is set to a number of Monte Carlo steps equal to the number of
vertices in the tiling and is called a Monte Carlo sweep (MCS). We
start with a tiling made by flat and equally spaced de Bruijn
surfaces. We equilibrate it during a time $\tau$ estimated at the end
of this subsection. Typically, for a tiling of size $p=20$ that we
present here, the equilibration time $\tau$ is of order $10^5$ MCS. As
we explained it above, we then choose a small central cluster of
vertices $i$, the position $r_i(t)$ of which we follow as a function
of time. The number of vertices in the central cluster is around $4000$
for $p=20$ whereas the tiling contains more than $10^5$ vertices. We
compute the mean square displacement averaged over all vertices and
all samples: $\langle (r(t)-r(0))^2 \rangle$.

\begin{figure}[h!]
\begin{center} 
\rotatebox{-90}{\resizebox{5.6cm}{!}{\includegraphics{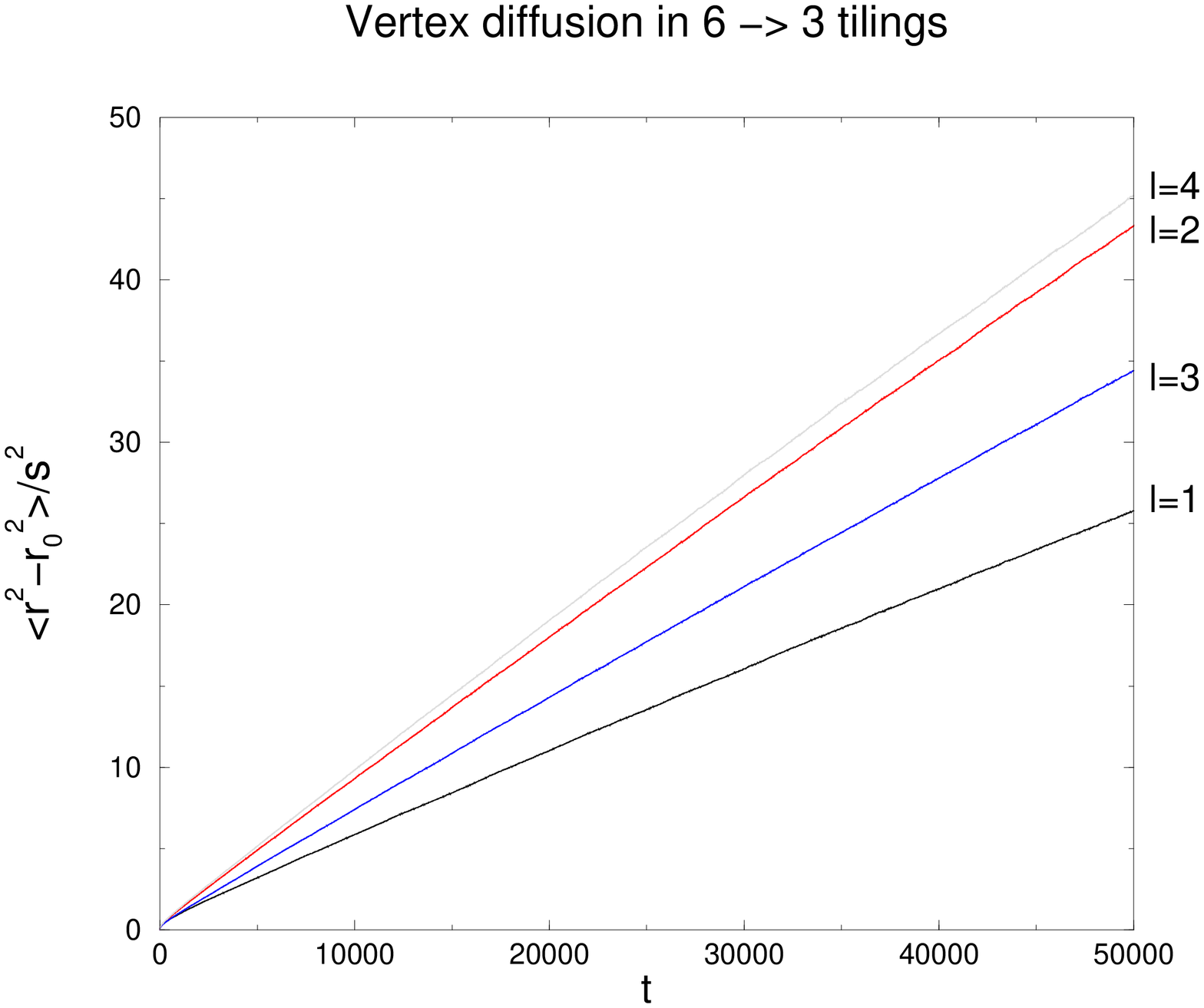}}}~\hfill~\rotatebox{-90}{\resizebox{5.6cm}{!}{\includegraphics{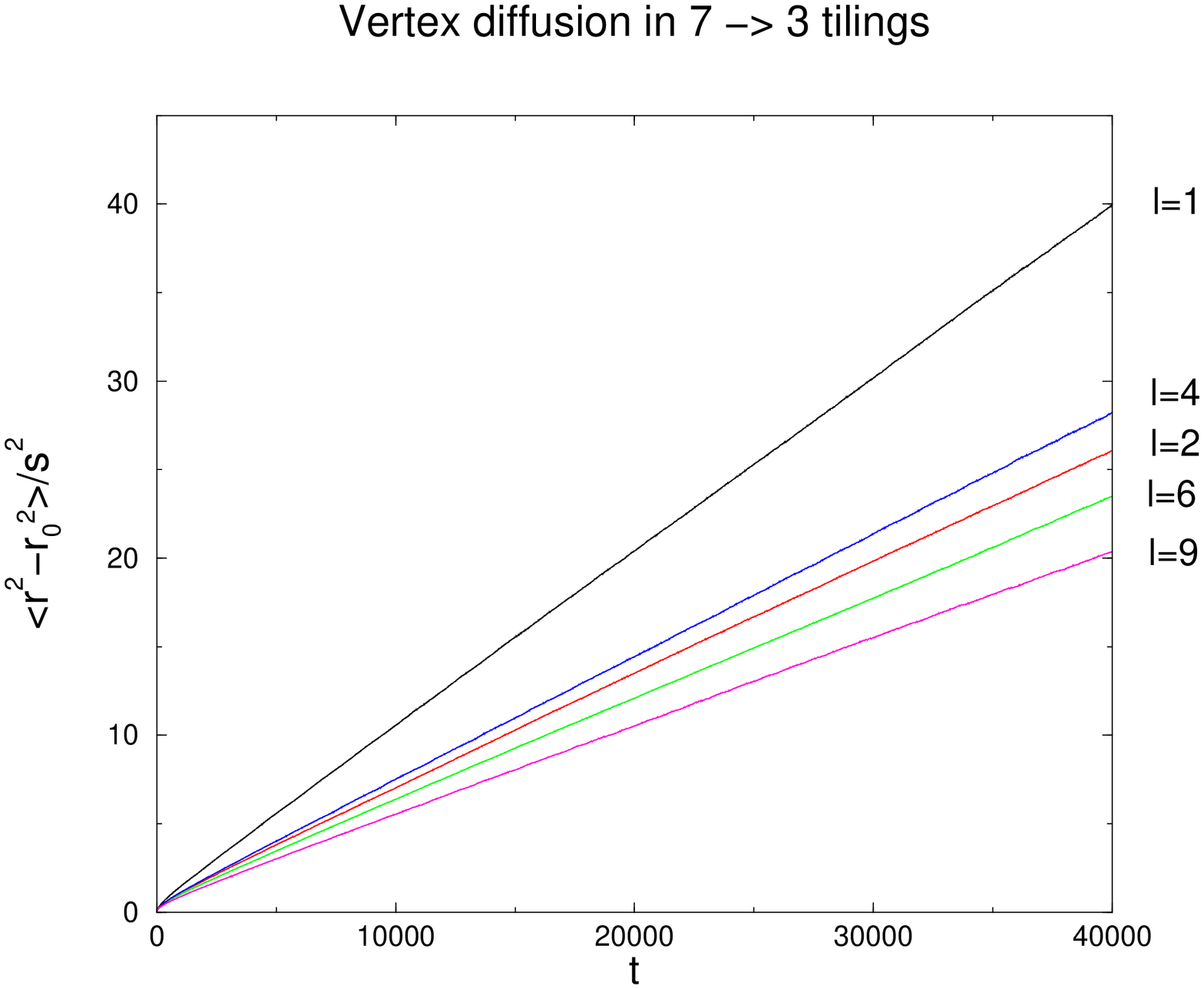}}}
\end{center}
\caption[]{Normalized mean square displacement of vertices in $6\ra3$
tilings (left) and $7\ra3$ tilings (right) with $p=20$. The $7\ra3$
line arrangements $l=1$ and $l=4$ correspond to tilings without cycles
in any fibrations, whereas the $l=2$ and $l=6$ ones correspond to
tilings with cycles in all fibrations.  The line arrangement $l=9$
corresponds to the intermediate case where cycles are present in some
fibrations but not all of them. For all line arrangements we find
normal diffusion after a transient.}
\label{f_diff63}
\end{figure}

As anticipated from reference~\cite{Jaric94}, this quantity grows like
$t$ at large time, indicating a diffusive regime, whatever the edge
orientation we choose, as displayed in figure~\ref{f_diff63}. We
comment these numerical results at the end of the section,
after clarifying some technical points.

First of all, the diffusion constant $\kappa$
depends not only on the codimension and on the equivalence class of
edge orientations, but also on the precise choice of these
orientations in a given equivalence class, since there is some liberty
of rotating and elongating the vectors $\evect{a}$ in a same
class. There is no obvious way in the general case of particularizing
a reference orientation in a given class. The only case where it is
possible is the first 6-line arrangement, for which edges pointing
towards the vertices of a regular icosahedron maximize the
symmetry. To smooth the differences between orientation vectors inside
an equivalence class, we normalize the mean square displacement by the
typical square distance $s^2=\langle \Delta r^2 \rangle$ covered by
the vertices at each step.

In addition, the mean square displacement exhibits a transition regime
before the diffusive one because of short-time correlations. This
regime stops around $\langle (r(t)-r(0))^2 \rangle/ s^2 \sim 1$. One
can interpret the duration of this transient regime as the typical
time $\tau_0$ between two uncorrelated flips~\cite{Jaric94}. Indeed, a
vertex just being flipped can only be flipped again to its initial
position at the next step. To have larger distances covered by a
vertex, the flip of this vertex has to be followed by a collective
sequence of vertex flips around it. During this sequence, the vertex
can go to a new flippable configuration. This collective succession of
flips should take a time of order $\tau_0$.  This time $\tau_0$ is
found to lie between $500$ and $1000$ MCS in all the cases studied
here. A vertex can be seen as a standard random walker making
independent steps of typical length $s$ every $\tau_0$ MCS and $\kappa
\approx s^2 / \tau_0$.

The diffusion constants we find after normalization do not depend much
on the codimension and the class of edge orientations, which indicates
that cycles have no clear influence on vertex diffusion. However, the
normalized diffusion constants we find for different edge orientations
in a same equivalent class show that this normalization is not
sufficient. The order of magnitude of the differences between these
diffusion constants in a same equivalence class is of the same order
as those between different classes. So we are not able to compare
quantitatively the differences in the diffusive dynamics between two
classes. However, we can display quantitative results in the case of
icosahedral symmetry where orientation vectors are well defined.  In
this case, we find normal diffusion with a {\em not} normalized
diffusion constant $\kappa = 0.0012$, which is very close to the one
found in reference~\cite{Jaric94} in the case of periodic-boundary
tilings.

But again, the aim of this study was more to observe the possible
fundamental differences in flip dynamics between tilings with and
without cycles. In particular to check if anomalous diffusion arises
in tilings with cycles. The results shown in figure~\ref{f_diff63}
present no anomalous diffusion whatever the tiling problem we are
considering. They also show that the diffusion constants are of the
same order of magnitude in the three cases: (i) no cycles in any
fibration; (ii) cycles in some fibrations but not all of them; (iii)
cycles in all fibrations (in which case the connectivity remains open).

As a conclusion, cycles do not seem to have any significant influence
on diffusive properties of vertices. We naturally expect that large
tilings inherit this diffusive behavior, and that vertices display the
same diffusive dynamics in free-boundary tilings related to real
quasicrystals, as it is argued in the previous section.

To close this section, we mention that the study of diffusion enables
a rough estimate of ergodic times in tiling sets. We recall that the
ergodic time $\tau$ of a Markovian process is the typical time the
process needs to reach stationarity, in other words to be likely to
have reached any configuration with nearly equal probability.  A time
scale can be associated with vertex diffusion, and is related to the
approach of stationarity: it is the typical time needed by a vertex
to explore the whole tiling, namely $\tau \sim p^2/\kappa$ for a tiling of
typical radius $p$. This time is compatible with known ergodic times
in dimension 2~\cite{Randall98,PRLbibi} and in $4 \ra 3$
tilings~\cite{Linde01}. Since $\kappa$ does not depend much
on the line arrangement, neither does this typical time $\tau$.

\section{Conclusion and discussion}
\label{cl}

This paper studies sets of three- (and higher-) dimensional tilings by
rhombohedra endowed with local rearrangements of tiles called
elementary flips or localized phasons. It uses a coding of tilings by
generalized partitions which turns out to be a powerful tool to prove
connectivity by flips in a large variety of cases. These results
answer positively (even though partially) an old conjecture of Las
Vergnas~\cite{LasVergnas80}. The general idea of the proof is as
follows: consider a tiling problem of codimension $c$. We intend to
prove the connectivity of its configuration space $G_c$. The
generalized partition-on-tiling point of view provides a natural
decomposition of $G$ into disjoint {\em fibers} above a {\em
base}. The base is the configuration space $G_{c-1}$ of a tiling
problem of codimension $c-1$. Therefore if it can be proven
inductively that the latter configuration space $G_{c-1}$ is connected
{\em and that all fibers are connected}, the overall connectivity of
$G_c$ follows.

So far, attempts of proofs of connectivity have failed because of the
possible existence of cycles in generalized partition problems. As it
is discussed in the paper, these cycles block locally some type of
flips, the proof of connectivity of fibers {\em a priori} fails and
the simple iterative proof of overall connectivity fails in its turn.
What we demonstrate in this paper is that this problem can be bypassed
in a large majority of cases because, in general, there exists one way
of implementing the generalized partition (one ``fibration'') so that
it does not generate cycles. As a consequence, fibers are connected.
Since it can also be proven inductively that the {\em base} is
connected, the overall connectivity can be established.

We say ``a large majority of cases'' because the result depends on the
choice of edge orientations. Indeed, we address in this paper the
implications in random tiling theory and quasicrystal science of this
issue. For example, beside the usual orientation of edges associated
with the icosahedral symmetry, there exist 3 additional non-equivalent
choices of edge orientations for codimension-three tilings. This means
that we consider tilings with the same number of different tile
species (namely 20), but the 4 sets of tile species are all non-equivalent
in that sense that the tilings they generate cannot be put in
one-to-one correspondence. It happens that cycles exist only in the
icosahedral case but cannot exist in the three remaining cases. As a
consequence, the only codimension-three case where we cannot establish
connectivity is the icosahedral one. Similarly, there are 11
non-equivalent edge orientations in codimension 4 and we prove
connectivity in all of them except 2. 

The counterparts of cycles at the tiling level are jammed clusters of
tiles that are more difficult to break by flips than the remainder of
the tiling because some type of flips is locally absent. It was
legitimate to anticipate that they might be responsible for entropic
barriers and slow down flip dynamics. We have chosen to address the
possible effects of cycles on flip dynamics from the angle of vertex
self-diffusion because it is a key issue at the physical level. We
prove in this paper that there exist tiling problems of the same
codimension with (i) no cycles in any fibration; (ii) cycles in some
fibrations but not all of them; (iii) cycles in all fibrations.
Connectivity holds in cases (i) and (ii) and remains open in case
(iii). We compared self-diffusion in the three cases, and we did not
detect any significant effect such as a sub-diffusive
regime. Therefore even if cycles break connectivity in case (iii),
they do not affect significantly the diffusive properties of physical
interest.

The tilings considered here have non-physical fixed boundaries.
However, we have argued that a macroscopic central region of tilings
is not influenced by the boundary and can therefore be considered as a
free-boundary one. This central region contains the jammed clusters of
tiles due to cycles. As a consequence, they are not specific to
fixed-boundary tilings. We concentrated our attention on this
region. We concluded that our results can be transposed to
free-boundary random tilings which are more realistic models of
quasicrystals.

Beyond diffusive properties, flip dynamics can be characterized 
by the calculation of ergodic times (the times needed to reach
stationarity in the flip Markovian process). We have not
addressed this question in the paper. The only conclusion that
we can draw is that a diffusive behavior is compatible with
ergodic times quadratic in the system size. This point will have
to be clarified in the future, but beyond numerical techniques,
the methods to tackle this point ought to be invented. The
standard methods in this field cannot easily be adapted because
of the existence of cycles which make impossible the calculation
of these times in fibers that are not connected.

Another issue that is not addressed in this paper is the influence of
energy interactions between tiles at finite temperature. More
realistic tiling models take into account a tile Hamiltonian
(reminiscent of interactions at the atomic level) that favors a
quasicrystalline order at low temperature.  We intend to analyze the
effects of cycles in this context in a future work.

To finish with, we mention that the possible influence of cycles can
be quantified on other observables than vertex self-diffusion. For
example, we have explored in a preliminary work how the parts
$X_u$ converge towards their average value. The part value in the
generalized partition formalism is an indication of the position of a
tile in the tiling. A deviation in the limiting values would mean that
some tiles do not find easily their equilibrium positions in the
tiling. It would be a manifestation of ergodicity (or even
connectivity) breaking. In general the convergence is rapid. However,
we have observed in rare circumstances in the case (iii) above that
these values do not converge exactly to their expected equilibrium
limit.  But in the state of progress of this work, it would be
premature to draw any conclusion because we are not able yet to
distinguish definitively between a real effect and statistical
noise. This work is in progress.

More generally, even if cycles do not perturb the physical properties
related to diffusion, we have not excluded the possible occurrence of
ergodicity (or even connectivity) breaking due to cycles in
icosahedral random tilings, which are related to real quasicrystals.
It could have important consequences on physical properties
related to flip dynamics but not directly to diffusion, such as
relaxation of the structure after a mechanical perturbation ({\em
e.g.} healing of phason faults behind a dislocation) or a quench.
Indeed even if it is always possible to break tiles in a
real quasicrystal to bypass a possible locking due to cycles, such a
process requires to pass an energy barrier, which might become
difficult at low temperature. Moreover, ergodicity
breaking would have dramatic consequences in Monte Carlo simulations
based on flip dynamics.

\section*{Acknowledgments}

One of us (ND) is indebted to Victor Reiner for making him aware of
the existence of cycles and of non-equivalent edge orientations in
dimension 3. We also express our gratitude to R\'emy Mosseri,
\'Eric R\'emila and Daniel Caillard for fruitful discussions and debates.

\appendix

\section{Connectivity of codimension 1 and 2 tiling sets in any dimension}
\label{codim2}

In this appendix, we prove connectivity of sets of codimension-1 and 2
rhombus tilings of any dimension $d$. In codimension 1, the proof is
immediate since such tilings are coded by (acyclic) hyper-solid
partitions~\cite{destain97} and since we prove in~\ref{connex_fib}
that they are consequently connected. In codimension 2, we also use a
proof by monotony as in section~\ref{connex}, even if we do not work
directly on hyper-plane arrangements in the projective space
$\Pb\Rb^{d-1}$.  Note that all edge orientations are equivalent in any
dimension and codimensions 1 and 2~\cite{remila04}.

We code $d+2 \ra d$ tilings as generalized partitions
on $d+1 \ra d$ codimension-one tilings. For sake of convenience, we
identify $\Rb^d$ with the hyperplane $H_d$ of $\Rb^{d+1}$ of equation
$\sum x_i =0$ and we choose the $d+1$ vectors $\evect{a}$ as follows:
\begin{eqnarray}
\evect{1} & = & (-d,1,\ldots,1), \nonumber \\
\evect{2} & = & (1,-d,1,\ldots,1), \nonumber \\
\vdots & \  \vdots & \ \ \ \vdots \\
\evect{d} & = & (1,\ldots,1,-d), \nonumber \\
\evect{d+1} & = & (-1,-1,\ldots,-1,d). \nonumber
\end{eqnarray}
The $(d+2)$-th orientation which orients base $d+1 \ra d$ tilings is
chosen as 
\begin{equation}
\evect{d+2} = (-1+\epsilon,-1+\epsilon^2,\ldots,
-1+\epsilon^d,d-(\epsilon+\ldots+\epsilon^d)), 
\end{equation}
where $\epsilon$ is a small positive parameter. This choice 
is a convenient one among any (non-degenerate) other one because all 
edge orientations are equivalent in codimension 2~\cite{remila04}.
A face of a base tiling $\tilde{t}$ is oriented accordingly to $\evect{d+2}$.
Now we exhibit precisely the orientation of each face species. 

A face species is defined by $d-1$ edge orientations among the $d+1$
possible ones. We denote by $a$ and $b$, $a < b$, the two indices of the
edge orientations that do {\em not} define a face species, and by
$\FC_{ab}$ this face species. We also denote by $\vect{g}_{ab}$
the vector normal to the faces $\FC_{ab}$. A simple calculation
shows that 
\begin{equation} 
\vect{g}_{ab} =(0,\ldots,0,1,0,\ldots,0,-1,0,\ldots,0),
\end{equation} 
where the non-zero coordinates are in positions $a$ and $b$.  We
define $\hat{\vect{g}}_{ab}=+\vect{g}_{ab}$ when $b \neq d+1$ and
$\hat{\vect{g}}_{ab}=-\vect{g}_{ab}$ when $b = d+1$.  Then
$\evect{d+2} \cdot \hat{\vect{g}}_{ab} > 0$ for any $a$ and $b$: a
face $\FC_{ab}$ is oriented positively in the direction
$\hat{\vect{g}}_{ab}$.

In addition, $\evect{d+1} \cdot \hat{\vect{g}}_{a,d+1} > 0$ whatever
$a<d+1$, which proves that $\evect{d+2}$ and $\evect{d+1}$ are
companion vectors, which in turn proves the monotony of the de Bruijn
indices $q_{d+1}$ with respect to the order between tiles. The
connectivity follows as in section~\ref{connex}.

\section{Connectivity of a fiber when the base tiling is acyclic}
\label{connex_fib}

In this section, we demonstrate that when a base tiling 
(or more generally a generalized partition problem) is
acyclic, the corresponding fiber is connected. We prove that
any partition $x$ can be connected to the minimum partition $z$
where all parts are set to 0.

We proceed by induction on the sum $\sigma(x)$ of the parts of
$x$. Suppose the result holds for all $x$ such that $ \sigma(x) \leq
\sigma_0$. Consider a partition $y$ with $ \sigma(y) =
\sigma_0+1$. 

Among all the parts of $y$ bearing non-zero parts, consider
a minimal one with respect to the order between parts. Such a
part exists because of the acyclic character of the partition problem.
Set this part to 0 by successive single flips. The so-obtained
partition $y'$ is connected to $z$ because $ \sigma(y') \leq
\sigma_0$, which proves that $y$ is connected by flips to $z$.

\end{document}